\documentclass[epjST]{svjour}

\usepackage[dvips]{graphicx}
\usepackage{bm}
\usepackage{amsmath}
\usepackage{amssymb}

\newcommand{\myand}{and }

\newcommand{\ovl}[1]{\overline{#1}}
\newcommand{\udl}[1]{\underline{#1}}

\newcommand{\dg}{\dagger}
\newcommand{\fdg}{{\phantom{\dagger}}}

\newcommand{\bvec}[1]{\vec{\bm #1}}

\newcommand{\id}{\mathrm{Id}}

\newcommand{\myskip}{\vspace{3mm}\noindent}

\newcommand{\be}{\begin{equation}}
\newcommand{\ee}{\end{equation}}

\newcommand{\ef}[1]{\, #1}

\newcommand{\diag}{\mathrm{diag}}

\newcommand{\cA}{\mathcal{A}}
\newcommand{\cL}{\mathcal{L}}
\newcommand{\cN}{\mathcal{N}}
\newcommand{\cM}{\mathcal{M}}

\newcommand{\relaxx}{\mathcal{R}}

\renewcommand{\emptyset}{\varnothing}
\renewcommand{\setminus}{\smallsetminus}

\newtheorem{theor}{Theorem} 
\newtheorem{coroll}{Corollary} 
\newtheorem{defin}{Definition} 
\newtheorem{prop}{Proposition} 
\def\proof{\par\medskip\noindent{\sc Proof.\ }}
\def\qed{\hfill $\square$}

\begin{document}


\title{Multiple and inverse topplings\\ in the Abelian Sandpile Model}

\author{Sergio Caracciolo
   \inst{1}\fnmsep\thanks{\email{s{}ergio.cara{}cciolo@m{}i.infn.it}} 
\and Guglielmo Paoletti
   \inst{2}\fnmsep\thanks{\email{pa{}oletti@df.un{}ipi.it}} 
\and Andrea Sportiello
   \inst{1}\fnmsep\thanks{\email{andre{}a.sportie{}llo@mi.in{}fn.it}} }
\institute{\inst{1} \rule{0pt}{10pt}Universit\`a degli Studi di
  Milano, Dipartimento di Fisica
      and INFN,\\via G.~Celoria 16, 20133 Milano, Italy \\
\inst{2} Universit\`a  di Pisa, Dipartimento di Fisica
      and INFN,\\largo B.~Pontecorvo 3, 56127 Pisa, Italy
}
\abstract{The Abelian Sandpile Model is a cellular automaton whose
  discrete dynamics reaches an out-of-equilibrium steady state
  resembling avalanches in piles of sand.  The fundamental moves
  defining the dynamics are encoded by the \emph{toppling rules}.  The
  transition monoid corresponding to this dynamics in the set of
  stable configurations is {\em abelian}, a property which seems at
  the basis of our understanding of the model.  By including also
  \emph{antitoppling rules}, we introduce and investigate a larger
  monoid, which is \emph{not abelian} anymore.  We prove a number of
  algebraic properties of this monoid, and describe their practical
  implications on the emerging structures of the model.}


\maketitle


\section{Introduction}
\label{sec:intro}

In 1987 Bak, Tang and Wiesenfeld~\cite{BTW} proposed a simple cellular
automaton model of sandpile growth (since now BTW model) as an example of self-organized
criticality (SOC). These systems should be characterized by a dynamics
which sometimes, in an apparently unpredictable way, shows bursts of
activity, avalanches in the sandpile, which eventually drive the
system into an out-of-equilibrium steady state. The state is a steady
state, in the sense that overall properties stay unchanged during time
evolution, and out-of-equilibrium because these systems are open and
dissipative and, therefore, they require input from outside at a
constant rate to balance the dissipation, and present a spatial
distribution of current. These steady states are critical as they
exhibit long-range correlations with power-law decay.

Consider a rectangular portion of a square lattice of size $L_x \times
L_y$, and let $n = L_x L_y$. To each site $i=(x,y)$ is associated a
non-negative integer \emph{height variable} (sometimes called
\emph{mass}) $z_i$.  A {\em configuration} is the collection of all
heights and will be denoted as a bold letter: $\vec{z}=\{ z_i
\}_{i=1}^n$.  If all $z_i$'s are smaller than $4$, the configuration
is \emph{stable}. Otherwise,
a procedure called  
\emph{relaxation} gives univocally a new configuration
$\relaxx(\vec{z})$. The relaxation is performed through a sequence of
elementary moves, called \emph{topplings}: if $z_{(x,y)}\geq 4$ (in
this case we say that the site $(x,y)$ is \emph{unstable}), then we
decrease $z_{(x,y)}$ by $4$, and increase by $1$ all the values at
neighbouring sites, $z_{(x\pm 1,y)}$ and $z_{(x,y\pm 1)}$ (nothing is
done at those coordinates $(x\pm 1,y)$, $(x,y\pm 1)$ falling outside
the grid). Topplings are performed at the various unstable sites,
until when the configuration obtained is stable.

It is remarkable, and important, that the final result does not depend
on the order in which the topplings are performed. In particular, if
$i$ and $j$ are unstable sites in the configuration $\vec{z}$, then $j$ is
still unstable after a toppling has been performed at $i$ (and
viceversa). Indeed, all the valid sequences of topplings differ only
for their ordering (although not all the permutations of a valid
sequence are valid sequences).

Note that, if we call $|\vec{z}| = \sum_{i=1}^n z_i$ the \emph{total mass}
of the configuration, the mass decreases for topplings occurring at the
boundary of the grid, and is conserved for topplings occurring in the
bulk. 

Call $a_i$ the operator which maps a stable configuration to a new one
by the addition of a grain of sand at the site $i$,  followed, if it becomes unstable, by a
relaxation. We have $n$ operators $a_i$'s, acting as a transition
monoid on the set of stable configurations.  As first pointed out by
Dhar~\cite{Dhar}, these operators \emph{commute}, i.e., for all
$\vec{z}$, the configurations $a_i a_j \vec{z}$ and $a_j a_i \vec{z}$
coincide. As a result, the transition monoid is \emph{abelian}.  Under
the Markov Chain obtained by acting with the $a_i$'s, chosen at
random, the sandpile is ergodic in a subset of all possible stable
configurations. According to the general notion used in the analysis
of Markov Chains, these configurations are called \emph{recurrent},
and they are characterized by the absence of \emph{forbidden
  sub-configurations} (FSC). All recurrent configurations occur with
uniform probabilities in the steady state of this dynamics. A stable
configuration which is not recurrent, i.e.\ that has zero steady-state
probability, is called \emph{transient}. As we will recall later on,
the notions of recurrent and transient configurations can also be
posed in purely algebraic terms, with no need to refer to
probabilistic notions.

In the steady state the fluxes of sand added to and flowing out of the
system must balance.
This balancing occurs however in a non-trivial way, as the
distribution of the size of the \emph{avalanches}, i.e., the number of
topplings required in the relaxation process at a given step of the
Markov Chain, has an algebraic tail, although the precise exponent is
still unknown~\cite{aval1,aval2,aval3,aval4}.

The algebraic
behaviour of the distribution of avalanche sizes suggests a relation
with a state at criticality.   
Indeed, there is an equivalence between the sandpile model
and the $q \to 0$ limit of the $q$-state Potts model, at criticality,
through a common combinatorial description in terms of (rooted)
spanning trees~\cite{Dhar2}. Therefore this sandpile critical behaviour
is in the same universality class of a conformal field theory (CFT) with
central charge $c=-2$, a relation which has allowed to obtain many
exact results on the critical exponents associated to quantities that,
differently from avalanche sizes, are more easily related to fields in
the CFT. For example, the correlation of height variables at two sites
decays only algebraically with the Euclidean distance between
them~\cite{MajDhar91}, the distribution of the heights has
corrections, depending on the Euclidean distance from the boundary,
that decay only algebraically,
and, consistently with the fact that the CFT at $c=-2$ is logarithmic,
certain two-point correlation functions, or one-point boundary
correlation functions, show the logarithmic
corrections to power-law decays that one expects from the presence of
Jordan blocks in the representation of the operator $L_0$ of the
Virasoro Algebra (see
\cite{priez,priez2,ruelle,ruelle2,jeng,ruelle3,ruepir_new}
for a collection of both rigorous and non-rigorous theoretical
results in this direction).

The dynamics of the sandpile, historically introduced first
in~\cite{BTW} for the square lattice, is so simple and natural that it
has been soon generalized to an arbitrary graph~\cite{Dhar}, an
extension that allows to elucidate the main algebraic elements of
the theory, and also independently studied by combinatorialists, under
the name of \emph{Chip Firing
  Game}~\cite{Lovasz,Biggs1,Biggs2,Lopez,Cori,Bernardi}, in relation
to the Tutte polynomial of the graph (not surprisingly, as the Tutte
polynomial of a graph is a reformulation of the Potts Model partition
function, the adjacency between variables being encoded by the graph,
see e.g.\ \cite{AlanBCC}). In particular, a series of papers
\cite{Lopez,Cori,Bernardi} shows how the generating function of
recurrent configurations, weighted according to their total mass, gives
access to larger class of enumerating problems, in particular,
of spanning connected subgraphs and spanning forests.

Also our personal interest was induced by the relation of the sandpile
model on an arbitrary graph and the $q \to 0$ Potts model,
because we have introduced a novel representation in terms of
fermionic variables of a model of random forests~\cite{Car04,Car07},
that generalizes the Kirchhoff Matrix-Tree Theorem, corresponding to
the case of spanning trees. 

More recently, we have been strongly impressed by some general
results which have been obtained for dynamics where all the sand is
deterministically injected in a limited region~\cite{Ostojic,DSC},
producing complex and beautiful patterns which display allometry. This
has led to a long-term research project, of which some first results
are reported in \cite{noi-epl}.  We believe that an important
ingredient to better understand the emergence of these patterns, even
for the ordinary Abelian Sandpile described above, is to understand
two distinct, and apparently unrelated, generalisations of the model.
The first more general toppling rules concern \emph{multitopplings},
i.e.\ topplings associated to clusters of sites~\cite{Gu,HLMPW}. The second one
concerns \emph{antitopplings}, where instead of adding sand, and then
relaxing, the sand is removed, and then an inverse relaxation is
performed. An analogous of this latter elementary move had already
been considered in~\cite{DharManna}, with the aim of determine
theoretically certain critical exponents associated to the model in
two dimensions, and, in the same variant considered here, has been
studied later on, mostly numerically, in~\cite{MannaPH}.

In this paper we will mostly present theoretical results for the
transition monoid associated to the sandpile in which both topplings
and antitopplings are allowed (remark that this monoid is no more
abelian). These results are in the form of non-trivial algebraic
relations (see e.g.\ Theorem \ref{theo:mainweak} later on).  Along the
paper we will see how one of the theorems in the theory of
antitopplings (Theorem \ref{theo.orchidweak}) has a relatively simple
proof only in the framework of the theory of multitopplings, this
explaining the need of treating the two generalisations
simultaneously.

The paper is organised as follows. Section \ref{sec:algdef}, by far
the longest of the paper, gives in Section \ref{ssec:basics} a first
reminder of the properties of the Abelian Sandpile Model, then
summarizes in Section \ref{ssec:results} a collection of new
theoretical results, finally in Section \ref{ssec:further} more subtle
aspects of the Abelian Sandpile Model are reviewed in detail.  Due to
a natural involution symmetry between toppling and antitopplings, the
latter are introduced from the very beginning, even in the review
parts. Conversely, the introduction and study of multitopplings is
postponed to Section \ref{sec:multitop}, where Theorem
\ref{theo.orchidweak}, presented in Section \ref{ssec:results}, is
proven. Section \ref{sec:dynams} presents numerical examples (on the
BTW setting) that motivate the study of several sandpile dynamics
involving simultaneously topplings and antitopplings.

\section{Algebraic formalism}
\label{sec:algdef}

The reader interested in the general theory of Abelian Sandpiles will
find in~\cite{Dhar99} a beautiful review. An updated version is
in~\cite{Dhar2006}, and other details are reported in~\cite{Redig}.
In Section \ref{ssec:basics} we shall first report the basic facts
that we need for our new developments.  Note that, differently from
the treatments in the references above, we introduce antitopplings, as
well as several other ``anti-'' quantities, from the very beginning,
and we choose a notation that highlights the role of a simple natural
involution, that exchanges ordinary and the conjugated
\hbox{``anti-''} quantities.

Facts already described in \cite{Dhar99} are reported here for both
classes of quantities.  Indeed, because of the involution symmetry,
the action of the conjugated operators alone cannot lead to any
special novelty. Nonetheless, non-trivial aspects will arise when
ordinary and conjugated operators \emph{simultaneously} enter the
game. The first simple properties in this wider context are discussed
along Section \ref{ssec:basics}, while the most remarkable facts are
anticipated in Section \ref{ssec:results}, and proven in the body of
the paper. Section \ref{ssec:further} presents further aspects of the
theory as developed in \cite{Dhar99}, that are not required
for the understating of the statements of our results, but are useful
in their proofs.

\subsection{Definitions and basic facts}
\label{ssec:basics}

Let $n$ be an integer, the \emph{size} of the system. It is often useful
to think at the system as a graph with $n$ sites, and the set of
toppling rules in terms of the adjacency structure of the graph, thus
we will call \emph{sites} the indices $i \in [n] \equiv
\{1,\ldots,n\}$.  Consider vectors $\vec{z} \in \mathbb{Z}^n$, where
we will use the partial ordering $\preceq$ such that $\vec{u} \preceq
\vec{v}$ if $u_i \leq v_i$ for each $i \in [n]$.  We will define the
\emph{positive cone} $\Omega$ as the subset of $\vec{w} \in
\mathbb{Z}^n$ such that $\vec{w} \succeq \vec{0}$, where $\vec{0}$ has
vanishing entries for all~$i$.

An abelian sandpile 
$\cA = \cA(\Delta, \ovl{\vec{z}}, \udl{\vec{z}})$
is identified by a triple $(\Delta, \ovl{\vec{z}}, \udl{\vec{z}})$,
that we now describe.
The vectors $\ovl{\vec{z}}$ and $\udl{\vec{z}}$ are
the collection of \emph{upper-} and \emph{lower-thresholds}, $\{
\ovl{z}_i \}$ and $\{ \udl{z}_i \}$ respectively, and are constrained
to the condition
$\ovl{z}_i - \udl{z}_i > 0$ for all $i$. Define the spaces
\begin{align}
S_+ &= \{\vec{z} \in \mathbb{Z}^n \,  
| \, \vec{z} \succeq \udl{\vec{z}} \} =
\textrm{{\scriptsize $\bigotimes_{i=1}^n$}} 
\{ \udl{z}_i, \udl{z}_i+1, \ldots \}
\ef;
\\
S_- &= \{\vec{z} \in \mathbb{Z}^n \,  
| \, \vec{z} \preceq \ovl{\vec{z}} \} =
\textrm{{\scriptsize $\bigotimes_{i=1}^n$}} 
\{ \ldots, \ovl{z}_i-1, \ovl{z}_i \}
\ef;
\\
S &= \{\vec{z} \in \mathbb{Z}^n \,  
| \, \udl{\vec{z}} \preceq \vec{z} \preceq \ovl{\vec{z}} \} =
\textrm{{\scriptsize $\bigotimes_{i=1}^n$}} 
\{ \udl{z}_i, \ldots, \ovl{z}_i \} = S_+ \cap S_-
\ef.
\end{align}
Thus $S_+$ and $-S_-$ are translations of $\Omega$, while $S$ is a
multidimensional interval, and has finite cardinality.

We also have a $n \times n$ \emph{toppling matrix} $\Delta$, with
integer entries, that should satisfy
$\ovl{z}_i - \udl{z}_i +1 \geq \Delta_{ii} > 0$, and $\Delta_{ij} \leq
0$ for $i \neq j$.  We say that the sandpile is \emph{tight} if
$\ovl{z}_i - \udl{z}_i +1 = \Delta_{ii}$ for all $i$.  Define the
spaces
\begin{align}
S'_+ &=
\textrm{{\scriptsize $\bigotimes_{i=1}^n$}} 
\{ \ovl{z}_i - \Delta_{ii}+1, \ovl{z}_i - \Delta_{ii}+2, \ldots \} 
\ef;
\\
S'_- &=
\textrm{{\scriptsize $\bigotimes_{i=1}^n$}} 
\{ \ldots, \udl{z}_i + \Delta_{ii} -2, \udl{z}_i + \Delta_{ii} -1 \}
\ef.
\end{align}
We have $S_+ \supseteq S'_+$, and $S_- \supseteq S'_-$ in general,
while $S_+ \equiv S'_+$, and $S_- \equiv S'_-$, if and only if the
sandpile is tight.

We further require \emph{dissipativity}, that is $b_i^- = \sum_j
\Delta_{ij} \geq 0$.  As seen below, $b_i^-$ is the amount of mass
that \emph{leaves} the system after a toppling at $i$. The requirement
that the toppling matrix is \emph{irreducible}\footnote{This means
  that for every $j_0$ there exists a sequence $(j_0, j_1, \ldots,
  j_{\ell})$ such that $\Delta_{j_a\,j_{a+1}} < 0$ for all $0 \leq a <
  \ell$, and $b^-_{j_{\ell}} > 0$.}  ensures that the avalanches are
finite (and that $\det \Delta>0$).  For future utility, we also
define $b_j^+ = \sum_i \Delta_{ij}$, 
which is the difference between the amount of mass that can leave the site $j$ in a toppling and the mass that can be added if all other sites would make a toppling.
A site $j$ where $b_j^+<0$ is said to be {\em greedy} or {\em selfish}.
Clearly  $\sum_i b^-_i = \sum_i b^+_i$, so also the $b^+_i$'s are, on average,
higher than zero, however positivity on the $b^+_i$'s is not implied
directly by the positivity of the $b^-_i$'s. We require the absence of greedy sites
as an extra condition, whose utility will be clear only in the
following. A sandpile is said to be \emph{unoriented} if
$\Delta =\Delta^{\rm T}$ (and thus $\vec{b}^- = \vec{b}^+$).

The above conditions
complete the list of constraints characterizing valid triples
$(\Delta, \ovl{\vec{z}}, \udl{\vec{z}})$.  The special case of the BTW
sandpile corresponds to $\Delta$ being the discretised Laplacian on
the square lattice, $\Delta_{ii}=4$, $\Delta_{ij}=-1$ if $d(i, j)=1$
and $\Delta_{ij} = 0$ if $d(i, j) >1$, where $d(i, j)$ is the
Euclidean distance between the sites $i$ and $j$, and $\ovl{z}_i=3$,
$\udl{z}_i=0$ for all $i$.

The matrix $\Delta$ is the collection of the toppling rules, and is
conveniently seen as a set of row vectors $\bvec{\Delta}_i = \{
\Delta_{ij} \}_{1 \leq j \leq n}$. Denote by $t_i$ the action of a
toppling at $i$. If such a toppling occurs, the configuration
$\vec{z}$ is transformed according to
\be
\label{eq.top}
t_i \vec{z} = \vec{z} - \bvec{\Delta}_i
\ef.
\ee
A site $i$ is \emph{positively-unstable} (or just \emph{unstable}) if
$z_i > \ovl{z}_i$. In this case, and only in this case, a toppling
can be performed at $i$. Note that, after the toppling, it is still
$z_i \geq \udl{z}_i$ (more precisely, $z_i > \ovl{z}_i -
\Delta_{ii}$), while $z_j$'s for $j \neq i$ have not decreased, thus
the topplings leave stable both spaces $S_+$ and $S_+'$.
The \emph{relaxation operator} $\relaxx$ is the map from $S_+$ to $S$,
coinciding with the identity on $S$, that associates to a
configuration $\vec{z} \in S_+$ the unique configuration
$\relaxx(\vec{z}) \in S$ resulting from the application of topplings
at unstable sites.
Unicity relies on the abelianity of the toppling rules, i.e.\ that, if
$i$ and $j$ are both unstable for $\vec{z}$, $j$ is unstable for $t_i
\vec{z}$, and thus crucially relies on the fact that $\Delta_{ij}\leq
0$ for $i \neq j$.

A site $i$ is \emph{negatively-unstable} if $z_i < \udl{z}_i$. In
this case, an \emph{antitoppling} can be performed at $i$, with the rule
\be
\label{eq.antitop}
t^\dg_i \vec{z} = \vec{z} + \bvec{\Delta}_i
\ef.
\ee
We use deliberately the symbol $t^\dg_i$ instead of $t_i^{-1}$
because, although the effect of the linear transformation
(\ref{eq.antitop}) is just the inverse of the effect of
(\ref{eq.top}), the \emph{if} conditions for applicability of the two
operators are different.

Now antitopplings leave stable both spaces $S_-$ and $S'_-$.
The \emph{antirelaxation operator} $\relaxx^\dg$ is
the map from $S_-$ to $S$, coinciding with the identity on $S$, that
associates to a configuration $\vec{z} \in S_-$ the unique
configuration $\relaxx^\dg(\vec{z}) \in S$ resulting from the
application of antitopplings at negatively-unstable sites. As a matter of fact,
the involution 
\be
\label{eq.iota}
\iota 
\ : \quad
\vec{z} \to \ovl{\vec{z}} + \udl{\vec{z}} - \vec{z}
\ee
exchanges the role of operators with and without the $\dg$ suffix,
i.e., for the operators above and all the others introduced later on,
we have $A^\dg(\vec{z}) \equiv \iota A(\iota \vec{z})$.

Note that, for a configuration $\vec{z} \in \mathbb{Z}^n$, we cannot
exchange in general the ordering of topplings and antitopplings (i.e.,
if $i$ and $j$ are respectively positively- and negatively-unstable
for $\vec{z}$, $j$ might be stable for $t_i \vec{z}$, and $i$ might be
stable for $t_j^\dg \vec{z}$). Consistently, we do \emph{not} define
any relaxation-like operator from $\mathbb{Z}^n$ to $S$, as it would
be intrinsically ambiguous, and even not finite, even when the
toppling matrix is dissipative~\footnote{For example, in a sandpile
  that has all $\ovl{z}=2$, all $\udl{z}=0$, all diagonal
  $\Delta_{aa}=3$, and non-zero $\Delta_{ab}$ with $a=i,j,k$ given by
  the list $\{\Delta_{ij},\Delta_{jk},\Delta_{ki},
  \Delta_{ii_{1,2}},\Delta_{jj_{1,2}},\Delta_{kk_{1,2}} \}$ and all
  equal to $-1$,
in the configuration $\vec{z}$ depicted below one can repeat
infinitely many times the cycle
$(
t^\dg_j
t^\fdg_k
t^\dg_i
t^\fdg_j
t^\dg_k
t^\fdg_i
) \vec{z}= \vec{z}$.
\[
\setlength{\unitlength}{12pt}
\raisebox{30pt}{$\cdots
  \stackrel{\displaystyle{t^\dg_j}}{\longrightarrow}$}
\!\!
\begin{picture}(8,7)
\put(2.9,4.25){\makebox[0pt][c]{-\hspace*{-0.5pt}$1$}}
\put(5.1,4.25){\makebox[0pt][c]{$2$}}
\put(7.3,4.25){\makebox[0pt][c]{$0$}}
\put(6.2,6.15){\makebox[0pt][c]{$0$}}
\put(0.7,4.25){\makebox[0pt][c]{$1$}}
\put(1.8,6.15){\makebox[0pt][c]{$1$}}
\put(4.0,2.35){\makebox[0pt][c]{$3$}}
\put(2.9,0.45){\makebox[0pt][c]{$0$}}
\put(5.1,0.45){\makebox[0pt][c]{$0$}}
\put(4.4,3.7){\scriptsize{$j$}}
\put(2.2,3.7){\scriptsize{$k$}}
\put(3.15,2.1){\scriptsize{$i$}}
\put(-0.02,-0.02){\includegraphics[scale=.6]{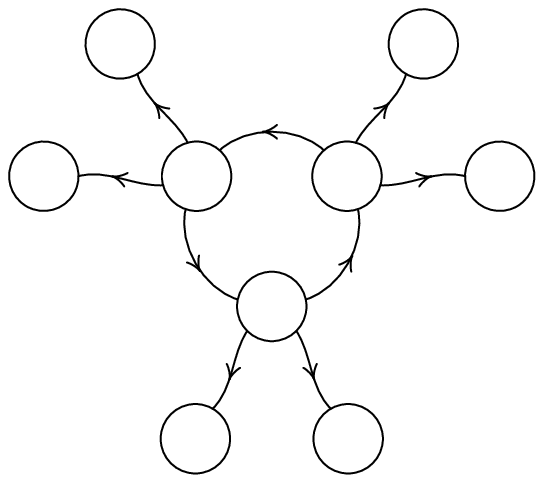}}
\end{picture}
\!\!
\raisebox{30pt}{$\stackrel{\displaystyle{t^\fdg_i}}{\longrightarrow}$}
\!\!
\begin{picture}(8,7)
\put(2.9,4.25){\makebox[0pt][c]{-\hspace*{-0.5pt}$1$}}
\put(5.1,4.25){\makebox[0pt][c]{$3$}}
\put(7.3,4.25){\makebox[0pt][c]{$0$}}
\put(6.2,6.15){\makebox[0pt][c]{$0$}}
\put(0.7,4.25){\makebox[0pt][c]{$1$}}
\put(1.8,6.15){\makebox[0pt][c]{$1$}}
\put(4.0,2.35){\makebox[0pt][c]{$0$}}
\put(2.9,0.45){\makebox[0pt][c]{$1$}}
\put(5.1,0.45){\makebox[0pt][c]{$1$}}
\put(4.4,3.7){\scriptsize{$j$}}
\put(2.2,3.7){\scriptsize{$k$}}
\put(3.15,2.1){\scriptsize{$i$}}
\put(-0.02,-0.02){\includegraphics[scale=.6]{fig_CicloInf.eps}}
\end{picture}
\!\!
\raisebox{30pt}{$\stackrel{\displaystyle{t^\dg_k}}{\longrightarrow}$}
\!\!
\begin{picture}(8,7)
\put(2.9,4.25){\makebox[0pt][c]{$2$}}
\put(5.1,4.25){\makebox[0pt][c]{$3$}}
\put(7.3,4.25){\makebox[0pt][c]{$0$}}
\put(6.2,6.15){\makebox[0pt][c]{$0$}}
\put(0.7,4.25){\makebox[0pt][c]{$0$}}
\put(1.8,6.15){\makebox[0pt][c]{$0$}}
\put(4.0,2.35){\makebox[0pt][c]{-\hspace*{-0.5pt}$1$}}
\put(2.9,0.45){\makebox[0pt][c]{$1$}}
\put(5.1,0.45){\makebox[0pt][c]{$1$}}
\put(4.4,3.7){\scriptsize{$j$}}
\put(2.2,3.7){\scriptsize{$k$}}
\put(3.15,2.1){\scriptsize{$i$}}
\put(-0.02,-0.02){\includegraphics[scale=.6]{fig_CicloInf.eps}}
\end{picture}
\!\!
\raisebox{30pt}{$\stackrel{\displaystyle{t^\fdg_j}}{\longrightarrow} \cdots$}
\]
}.

Remark, however, that the definition of $\relaxx$ can be trivially extended in order to map
unambiguously $\mathbb{Z}^n$ to $S_-$, by letting it produce a toppling only on unstable sites, and, analogously, $\relaxx^\dg$ to
$\mathbb{Z}^n$ to $S_+$ (and both $\relaxx^\dg \relaxx$ and $\relaxx
\relaxx^\dg$ map $\mathbb{Z}^n$ to $S$, but 
$\relaxx^\dg (\relaxx (\vec{z})) \neq \relaxx (\relaxx^\dg (\vec{z}))$
in general).


On the space $\mathbb{Z}^n$ we can of course take linear combinations,
and define the \emph{sum} of two configurations, $\vec{z} + \vec{w}$,
and the \emph{multiplication by scalars} $k \in \mathbb{Z}$,
$k\vec{z}$. For generic $\ovl{\vec{z}}$ and $\udl{\vec{z}}$ these
operations do not leave stable any of the subsets $S_{\pm}$ and
$S$. But the sum and difference, $\vec{z} + \vec{w}$ and $\vec{z} -
\vec{w}$, are binary operations in the following spaces:
\begin{align}
\vec{z} + \vec{w} :&&
S_+ \times \Omega &\to S_+
\ef;
&
S'_+ \times \Omega &\to S'_+
\ef;
\\
\vec{z} - \vec{w} :&&
S_- \times \Omega &\to S_-
\ef;
&
S'_- \times \Omega &\to S'_-
\ef.
\end{align}
Thus, in particular, the two maps $\relaxx(\vec{z} + \vec{w})$ and
$\relaxx^\dg(\vec{z} - \vec{w})$ can act from $S \times \Omega$ into
$S$. This suggests to give a special name and symbol to the simplest
family of these binary operations, seen as operators on $S$.  Call
$\vec{e}_i$ the canonical basis of $\mathbb{Z}^n$,
i.e.\ $(\vec{e}_i)_j = \delta_{i,j}$. Define the operators $\hat{a}_i$
such that $\hat{a}_i \vec{z} = \vec{z} + \vec{e}_i$, and introduce the
operators of
\emph{sand addition} and \emph{removal}
\begin{align}
a_i &=
\relaxx \hat{a}_i
\ef;
&
a_i^\dg &=
\relaxx^\dg \hat{a}_i^{-1}
\ef.
\end{align}
The $a_i$'s commute among themselves, i.e., for every $\vec{z}$,
$a_i a_j \vec{z} = a_j a_i \vec{z} =
\relaxx(\vec{z}+\vec{e}_i+\vec{e}_j)$. Similarly,
the $a_i^\dg$'s commute among themselves.  More generally, for
any $\vec{z} \in \mathbb{Z}^n$ and $\vec{w} \in \Omega$,
$\relaxx(\relaxx(\vec{z}) + \vec{w}) = \relaxx(\vec{z} + \vec{w})$,
and $\relaxx^\dg(\relaxx^\dg(\vec{z}) -\vec{w}) =
\relaxx^\dg(\vec{z} - \vec{w})$, this implying
\begin{align}
\relaxx(\vec{z}+\vec{w})
&=
\Big( \prod_{i=1}^n (a_i)^{w_i} \Big) \, \vec{z}
\ef;
&
\relaxx^\dg(\vec{z}-\vec{w})
&=
\Big( \prod_{i=1}^n (a_i^\dg)^{w_i} \Big) \, \vec{z}
\ef.
\end{align}
This can be seen by induction in $i$, as 
$\relaxx(\vec{z}+(\vec{w}'+\vec{e}_i))=
\relaxx(\relaxx(\vec{z}+\vec{e}_i)+\vec{w}')
=\relaxx((a_i \vec{z})+\vec{w}')$).
For later convenience,
for
$\vec{w} \in \Omega$, we introduce the shortcuts
\begin{align}
a_{\vec{w}}^\fdg 
&=
\prod_{i=1}^n (a_i)^{w_i} = \relaxx(\,\cdot\, + \vec{w})
\ef;
&
a^\dg_{\vec{w}} 
&= \prod_{i=1}^n (a_i^\dg)^{w_i} = \relaxx^\dg(\,\cdot\, - \vec{w})
\ef;
\end{align}
(note that the order in the products does not matter).

These properties have an important consequence on the structure of
the Markov Chain dynamics introduced in Section
\ref{sec:intro}: at each integer time $t$ a site $i(t)$ is chosen at random,
and $\vec{z}(t+1) = a_{i(t)} \vec{z}(t)$. As long as we are interested
in the configuration $\vec{z}(t_{\rm fin})$ for a unique final time
$t_{\rm fin}$, for a given initial state $\vec{z}(0)$, it is not
necessary to follow the entire evolution $\vec{z}(t)$, for
$0 \leq t \leq t_{\rm fin}$, but it is enough to take the vector
$\vec{w} = \sum_{t} \vec{e}_{i(t)}$, and evaluate $\relaxx(\vec{z}(0)
+ \vec{w})$. Thus, the final result of the time evolution depends on
the set
$\{ i(t) \}_{0 \leq t < t_{\rm max}}$ of moves at all times, in a
way which is invariant under their permutations. 

Note however that, similarly to topplings with antitopplings, also each
operator of sand addition and removal do not commute, 
i.e.\ $a_i^\fdg a_j^\dg \vec{z} \neq a_j^\dg a_i^\fdg
\vec{z}$ in general, even for $\vec{z} \in S$. Therefore, in a Markov
process involving both $a_i$'s and $a^\dg_i$'s, in order to know the
final configuration, it is necessary to follow the full trace of the
time evolution.  We will describe and briefly investigate a collection
of dynamics of this kind, for the sandpile on a square lattice, in
Section \ref{sec:dynams}. Furthermore, see \cite{MannaPH} for a first
extensive investigation of a dynamics in this family.

Beside the commutativity relations $a_i a_j = a_j a_i$ (and conjugated
ones), there is a collection of $n$ relations, encoded by the toppling
matrix: for any $i$, when acting on configurations
such that $z_i > \ovl{z}_i-\Delta_{ii}$, (respectively, such that $z_i
< \udl{z}_i+\Delta_{ii}$), we have
\begin{align}
\label{eq.ADeltaRela}
a_i^{\Delta_{ii}}
&=
\prod_{j \neq i} a_j^{-\Delta_{ij}}
\ef;
&
(a^\dg_i)^{\Delta_{ii}}
&=
\prod_{j \neq i} (a^\dg_j)^{-\Delta_{ij}}
\ef.
\end{align}
%
This because $t_i \hat{a}_i^{\Delta_{ii}} = \prod_{j \neq i}
\hat{a}_j^{-\Delta_{ij}}$ on such configurations, and the site $i$ is
certainly positively-unstable after the application of
$\hat{a}_i^{\Delta_{ii}}$.  As a corollary, relations valid for
generic configurations in $S_+$, and in $S_-$, respectively, are
\begin{align}
\label{eq.6543756b}
a_i^{\ovl{z}_i-\udl{z}_i+1}
&=
a_i^{\ovl{z}_i-\udl{z}_i+1 - \Delta_{ii}}
\prod_{j \neq i} a_j^{-\Delta_{ij}}
\ef;
\\
\label{eq.6543756c}
(a_i^\dg)^{\ovl{z}_i-\udl{z}_i+1}
&=
(a_i^\dg)^{\ovl{z}_i-\udl{z}_i+1 - \Delta_{ii}}
\prod_{j \neq i} (a_j^\dg)^{-\Delta_{ij}}
\ef;
\end{align}
which are concisely written as
\begin{align}
\label{eq.6543756bx}
a_i^{\ovl{z}_i-\udl{z}_i+1}
&=
a^\fdg_{-\vec{\Delta}_i + (\ovl{z}_i-\udl{z}_i+1) \vec{e}_i}
\ef;
\\
\label{eq.6543756cx}
(a_i^\dg)^{\ovl{z}_i-\udl{z}_i+1}
&=
a^\dg_{-\vec{\Delta}_i + (\ovl{z}_i-\udl{z}_i+1) \vec{e}_i}
\ef.
\end{align}
Remark that the two sets of equations (\ref{eq.ADeltaRela}) and
(\ref{eq.6543756b}) do coincide if the sandpile is tight, i.e.\ if
$\ovl{z}_i-\udl{z}_i+1 = \Delta_{ii}$ for all $i$.

\myskip
In the previous paragraphs we only considered sums $\vec{z} +
\vec{w}$, in which one of the two configurations is taken from a space
among $S$, $S_{\pm}$, or $S'_{\pm}$, and the other one
from $\Omega$, or other spaces with no dependence from $\ovl{\vec{z}}$
and $\udl{\vec{z}}$. 

In such a situation, we have a clear covariance of the notations under
an overall translation of the coordinates. I.e., for any $\vec{r}\in
\mathbb{Z}^n$, under the map $\vec{z} \to \vec{z} + \vec{r}$, we have
an isomorphism between the sandpile $\cA(\Delta, \ovl{\vec{z}},
\udl{\vec{z}})$ and $\cA(\Delta, \ovl{\vec{z}}+ \vec{r},
\udl{\vec{z}}+ \vec{r})$. We call a \emph{gauge invariance} of the
model the covariance explicitated above, and a \emph{gauge fixing} any
special choice of offset vector $\vec{r}$.

As always, a gauge fixing reduces the apparent number of parameters in
the model. Some special choices simplify the notations in certain
contexts. For example, we can set $\udl{\vec{z}} = \vec{0}$, so that
$S_+ \equiv \Omega$, that we call the 
\emph{$\udl{\vec{z}} = \vec{0}$ gauge}, or $\ovl{z}_i =
\Delta_{ii}-1$, so that $S'_+ \equiv \Omega$, that we call the
\emph{positive-cone gauge}.

In particular, the positive-cone gauge is the most natural one in the
Abelian Sandpile in which antitoppling are not considered, as in this
case the parameters $\udl{\vec{z}}$ do not play any role. Conversely,
the covariant formalism is the only formulation that does not break
explicitly the involution symmetry implied by (\ref{eq.iota}).

\subsection{Statement of results}
\label{ssec:results}

Two new theoretical results are extensively required for the analysis
of interesting dynamics, in Section \ref{sec:dynams}.
The first one is the following
\begin{theor}
\label{theo:mainweak}
For every $i \in [n]$, acting on $S_+$,
\be
\label{eq.intheo1w}
a_i^\fdg
a^\dg_i
a_i^\fdg
=
a_i^\fdg
\ef;
\ee
and, acting on $S_-$,
\be
\label{eq.intheo2w}
a^\dg_i
a_i^\fdg
a^\dg_i
=
a^\dg_i
\ef.
\ee
\end{theor}

\myskip

This theorem will be proven in the next subsection.
As an aside, we know that Theorem~\ref{theo:mainweak} is the tip of the
iceberg of a much more general family of identities, with operators
$a_i^\fdg$ and $a^\dg_i$ replaced by $a_{\vec{w}}^\fdg$ and
$a^\dg_{\vec{w}}$, for $\vec{w} \in \Omega$, that we plan to elucidate
in forthcoming works. Note for example how, acting on $S_+$, the
identity $a_{\vec{w}}^\fdg a^\dg_{\vec{w}} a_{\vec{w}}^\fdg =
a_{\vec{w}}^\fdg$ holds for simple reasons if $\vec{w}$ is
\emph{recurrent}, due to the fact that $a_{\vec{w}}^\fdg
a^\dg_{\vec{w}} a_{\vec{w}}^\fdg \vec{z} \sim a_{\vec{w}}^\fdg
\vec{z}$, both sides of the equation are stable recurrent, and there
exists a unique stable recurrent representative in each equivalence
class (see later Section \ref{ssec:further} for the pertinent
definitions).

An immediate corollary of Theorem~\ref{theo:mainweak} is the following
\begin{coroll}
\label{coro:idempweak}
For every $i \in [n]$, acting on $S$,
$a^\dg_i a_i^\fdg$ and
$a_i^\fdg a^\dg_i$ are
idempotents, i.e., 
$(a^\dg_i a_i^\fdg)^2 
= a^\dg_i a_i^\fdg$, and
$(a_i^\fdg a^\dg_i)^2
= a_i^\fdg a^\dg_i$.
\end{coroll}
Indeed, the set $S=S_+ \cap S_-$ is left stable by the action of the
monoid. It is enough to multiply equation (\ref{eq.intheo1w}) by
$a^\dg_i$, from the left or from the right respectively for the two
claims, or, alternatively, multiply (\ref{eq.intheo2w}) by $a_i^\fdg$,
from the right or from the left respectively.

The simplicity of Corollary \ref{coro:idempweak} may suggest that
abelianity is restored at the level of these idempotent
combinations. This is not the case. No pairs of distinct operators in
the set 
$\{ a^\dg_{1} a_{1}^\fdg, \ldots, a^\dg_{n} a_{n}^\fdg, 
a^\fdg_{1} a_{1}^\dg, \ldots, a^\fdg_{n} a_{n}^\dg \}$
commute with each other, in general. Nonetheless, a few interesting
facts are found.

For a finite set $I \subseteq [n]$, call $\cN_I = \{ a^\dg_i a^\fdg_i
\}_{i \in I}$.  For $X \subseteq \mathbb{Z}^n$, call $\cN_I[X]$ the
set of $\vec{y} \in \mathbb{Z}^n$ such there exists a configuration
$\vec{x} \in X$ and a finite sequence $(i_1,\ldots,i_k)$ of elements
in $I$ such that $a^\dg_{i_k} a_{i_k}^\fdg a^\dg_{i_{k-1}}
a_{i_{k-1}}^\fdg \cdots a^\dg_{i_1} a_{i_1}^\fdg \vec{x} = \vec{y}$,
that is $\cN_I[X]$ is the set of possible images of $X$ under the action 
of products of operators in $\cN_I$.

The second new theoretical result of this paper is
\begin{theor}
\label{theo.orchidweak}
Consider a sandpile $\cA(\Delta,\ovl{\vec{z}},\udl{\vec{z}})$ such
that $\Delta = \Delta^{\rm T}$ and $\Delta_{ij} \in \{0,-1\}$ for $i
\neq j$.  For any $I \subseteq [n]$, and any $\vec{z} \in S_+$, there
exists a unique state $\vec{y}(\vec{z},I)$ in $\cN_I[\{\vec{z}\}]$
such that $a^\dg_i a^\fdg_i \vec{y} = \vec{y}$ for all $i \in I$. For
any state $\vec{x} \in \cN_I[\{\vec{z}\}]$, we also have $\vec{y} \in
\cN_I[\{\vec{x}\}]$.
\end{theor}
Thus, this theorem shows that certain collections of idempotents have
a well-charac\-terised set of common fixed points. The portion of this
set accessible from any configuration $\vec{z}$ has cardinality
exactly~1.


The statement of this theorem can be translated in terms
of a stochastic evolution. Consider a Markov Chain in which the initial state
is $\vec{z}(0) = \vec{z}$, and at each time $t$ an element
$a^\dg_{i_t} a_{i_t}^\fdg \in \cN_I$ is chosen (with non-zero
probabilities for all elements), and $\vec{z}(t+1) = a^\dg_{i_t}
a_{i_t}^\fdg \vec{z}(t)$. Then the theorem states that this Markov
Chain is absorbent, on an unique state $\vec{y}$, and in particular,
no matter the evolution up to some time $t$, $\vec{y}$ is still
accessible from $\vec{z}(t)$ (and in fact it \emph{will} be reached at
some time). A dynamics of this kind will be described in
Section~\ref{ssec:dynproj}.


The theorem above will be proven in Section
\ref{sec:multitop}. Furthermore, the state $\vec{y}(\vec{z},I)$ will
be shown to have a further characterization, in terms of a
multitoppling Abelian Sandpile associated to the original system.

\subsection{Further aspects of the theory}
\label{ssec:further}

There exists a natural equivalence relation on vectors $\vec{z} \in
\mathbb{Z}^n$, that partitions this set into $\det \Delta$ classes,
which are affine subspaces of $\mathbb{Z}^n$, all isomorphic under
translation. This notion was first introduced and studied in
\cite{Dhar_alg}. We recall here only briefly the easiest facts.

We say that $\vec{z} \sim \vec{w}$ if there exists $\vec{T} \in
\mathbb{Z}^n$ such that $\vec{z} - \vec{w} = \vec{T} \Delta$.  In
particular, as $\vec{z} - t_i \vec{z} = - (\vec{z} - t_i^\dg \vec{z})
= \bvec{\Delta}_i = \vec{e}_i \Delta$, we have $\vec{z} \sim
\relaxx(\vec{z}) \sim \relaxx^\dg(\vec{z})$. Analogously, as
$\hat{a}_i^{\Delta_{ii}} \vec{z} - \big( \prod_{j \neq i}
\hat{a}_j^{-\Delta_{ij}} \big) \, \vec{z} = -\vec{e}_i \Delta$, we get
$ a_i^{\Delta_{ii}} \vec{z} \sim \big( \prod_{j \neq i}
a_j^{-\Delta_{ij}} \big)\,\vec{z} $ for all $i$, as it should at the
light of (\ref{eq.ADeltaRela}).  The dissipativity condition on
the toppling matrix ensure that $\det \Delta \neq 0$. Thus
$\Delta^{-1}$ exists, and it is evident that $\vec{z} \sim \vec{w}$ if
and only if $\vec{z} \Delta^{-1} - \vec{w} \Delta^{-1} \in
\mathbb{Z}^n$. So, the fractional parts $Q_i^{\rm (frac)}(\vec{z}) =
(\vec{z} \Delta^{-1})_i - \lfloor (\vec{z} \Delta^{-1})_i \rfloor$,
called the \emph{charges} of the configuration $\vec{z}$, completely
identify the equivalence class of $\vec{z}$. As a corollary,
$Q_i^{\rm (frac)}(\vec{z}) = Q_i^{\rm (frac)}(t^\fdg_i\vec{z}) 
= Q_i^{\rm (frac)}(t^\dg_i \vec{z})$ (when $t^\fdg_i$ or $t^\dg_i$ are
applicable to $\vec{z}$). For future convenience,
we also define $\vec{Q}(\vec{z}) = (\vec{z} \Delta^{-1})$, so that
$Q_i^{\rm (frac)} = Q_i - \lfloor Q_i \rfloor$~\footnote{This topic
  would deserve some more lines. The charges $Q_i^{\rm
    (frac)}(\vec{z})$ have two disadvantages: they are redundant in
  general, and require ``fractional parts'', instead of integer
  arithmetics. Both these issues are solved through the identification
  of a different set of charges, $I_i(\vec{z})$. A result of the
  classical theory of the Smith Normal Form is that any non-singular
  $n \times n$ matrix $\Delta$ can be decomposed as $\Delta = A D B$,
  where $A$ and $B$ are square matrices with determinant $\pm 1$, and
  $D=\diag(d_i)$, with $d_1 \geq d_2 \geq \cdots d_n$, all matrices
  having integer coefficients. The decomposition is not unique, but
  the $d_i$'s are, and have a further characterization that makes
  unicity manifest. Call $g_i$ the greatest common divisor
  among the minors of $\Delta$ of size $n-i$, and pose $g_n = 1$. Then
  $d_i=g_{i-1}/g_i$. One defines $I^{(0)}_i(\vec{z}) = (A^{-1}
  \vec{z})_i$, and $0 \leq I_i < d_i$, with $I_i \equiv I^{(0)}_i$ modulo
  $d_i$. Dropping the trivial charges, associated to $d_i=1$, the new
  set of charges is non-redundant and completely identifies the
  equivalence classes. The new charges are related to the previous
  ones through $\vec{I} = D B \,\vec{Q}$. A more detailed treatment can
  be found in the original reference,~\cite{Dhar_alg}. We do not make
  use of these improved charges in the present paper.}.

\myskip
The set $S$ of stable configurations is divided into the two subsets
of \emph{stable transient} and \emph{stable recurrent} configurations,
$S=T \cup R$. Several equivalent characterizations of recurrency
exist, some of which extend naturally to $S_+$, and even to the full
space $\mathbb{Z}^n$ (this is also our choice).




In particular, we give three definitions, all valid in $\mathbb{Z}^n$.
Under all definitions, a configuration is \emph{transient} if it is
not recurrent.  

\begin{defin}
A configuration $\vec{z}$ is \emph{recurrent by identity test} if
there exists a permutation $\sigma \in \mathfrak{S}_n$ such that 
$t_{\sigma(n)} \cdots t_{\sigma(2)} t_{\sigma(1)} (\vec{z} +
\vec{b}^+) = \vec{z}$ is a valid toppling sequence.
\end{defin}

\begin{defin}
A configuration $\vec{z}$ is \emph{recurrent by toppling covering} if
there exists a configuration $\vec{u}$, 
such that
$\vec{z} = t_{i_k} \cdots t_{i_1} \vec{u}$ is a valid toppling
sequence, and at least one toppling is performed at each site.
\end{defin}

\begin{defin}
A configuration $\vec{z}$ is \emph{recurrent by absence of FSC's} if,
for every set $I$ of sites, there exists $i \in I$ such that
$z_i > \ovl{z}_i - \sum_{j \in I} \Delta_{ji}$.
\end{defin}

\noindent
The reasonings of the following paragraphs will prove, among other
things, that these three definitions are equivalent.

First of all, a configuration recurrent by identity test is also
recurrent by toppling covering (one can take $\vec{u} = \vec{z} +
\vec{b}^+$).

Note that, if $\vec{u} = \relaxx(\vec{v})$ and $\vec{v}$ is recurrent
by toppling covering, also $\vec{u}$ is recurrent by toppling
covering.  Furthermore, as after a toppling $t_i$ one has $z_i >
\ovl{z}_i-\Delta_{ii}$, all the recurrent configurations, either by
toppling covering or by identity test, are in fact contained in
$S'_+$.  This last reasoning is a first example of 
\emph{forbidden sub-configuration} (FSC), whose generalisation
involves more than one site at a time.  For $\vec{z} \in
\mathbb{Z}^n$, and $I \subseteq [n]$, define
$\vec{z}|_I$ as the restriction to the components $z_i$ with index $i
\in I$.


For a set $I$, define the vector $\vec{f}_{\rm max}(I) \in
\mathbb{Z}^I$ as
\be
\label{eq.fmax}
\big(f_{\rm max}(I)\big)_i
=
\ovl{z}_i
-
\sum_{j \in I}
\Delta_{ji}
\ef.
\ee
We say that the pair $(I,\vec{z}|_I)$ is a 
\emph{forbidden sub-configuration} for $\vec{z}$ if 
$\vec{z} \preceq \vec{f}_{\rm max}(I)$.
It is straightforward to recognize that $\vec{z}$ is recurrent for
absence of FSC's if and only if, for all $I$, $\vec{z}|_I \not\preceq
\vec{f}_{\rm max}(I)$, thus legitimating the terminology.

A collection of the pairs $(I,\vec{f}_{\rm max}(I))$
with smallest $|I|$ in the BTW sandpile is as follows:
\[
\setlength{\unitlength}{12.5pt}
\begin{picture}(1,1)
\put(0,0){\line(1,0){1}}
\put(0,1){\line(1,0){1}}
\put(0,0){\line(0,1){1}}
\put(1,0){\line(0,1){1}}
\put(0.5,0.3){\makebox[0pt][c]{\scriptsize{$-1$}}}
\end{picture}
\qquad
\begin{picture}(2,1)
\put(0,0){\line(1,0){2}}
\put(0,1){\line(1,0){2}}
\put(0,0){\line(0,1){1}}
\put(1,0){\line(0,1){1}}
\put(2,0){\line(0,1){1}}
\put(0.5,0.25){\makebox[0pt][c]{$0$}}
\put(1.5,0.25){\makebox[0pt][c]{$0$}}
\end{picture}
\qquad
\begin{picture}(1,2)
\put(0,0){\line(1,0){1}}
\put(0,1){\line(1,0){1}}
\put(0,2){\line(1,0){1}}
\put(0,0){\line(0,1){2}}
\put(1,0){\line(0,1){2}}
\put(0.5,0.25){\makebox[0pt][c]{$0$}}
\put(0.5,1.25){\makebox[0pt][c]{$0$}}
\end{picture}
\qquad
\begin{picture}(3,1)
\put(0,0){\line(1,0){3}}
\put(0,1){\line(1,0){3}}
\put(0,0){\line(0,1){1}}
\put(1,0){\line(0,1){1}}
\put(2,0){\line(0,1){1}}
\put(3,0){\line(0,1){1}}
\put(0.5,0.25){\makebox[0pt][c]{$0$}}
\put(1.5,0.25){\makebox[0pt][c]{$1$}}
\put(2.5,0.25){\makebox[0pt][c]{$0$}}
\end{picture}
\qquad
\begin{picture}(2,2)
\put(0,0){\line(1,0){2}}
\put(0,1){\line(1,0){2}}
\put(0,2){\line(1,0){1}}
\put(0,0){\line(0,1){2}}
\put(1,0){\line(0,1){2}}
\put(2,0){\line(0,1){1}}
\put(0.5,0.25){\makebox[0pt][c]{$1$}}
\put(0.5,1.25){\makebox[0pt][c]{$0$}}
\put(1.5,0.25){\makebox[0pt][c]{$0$}}
\end{picture}
\quad
\cdots
\quad
\begin{picture}(4,1)
\put(0,0){\line(1,0){4}}
\put(0,1){\line(1,0){4}}
\put(0,0){\line(0,1){1}}
\put(1,0){\line(0,1){1}}
\put(2,0){\line(0,1){1}}
\put(3,0){\line(0,1){1}}
\put(4,0){\line(0,1){1}}
\put(0.5,0.25){\makebox[0pt][c]{$0$}}
\put(1.5,0.25){\makebox[0pt][c]{$1$}}
\put(2.5,0.25){\makebox[0pt][c]{$1$}}
\put(3.5,0.25){\makebox[0pt][c]{$0$}}
\end{picture}
\quad
\cdots
\quad
\begin{picture}(2,2)
\put(0,0){\line(1,0){2}}
\put(0,1){\line(1,0){2}}
\put(0,2){\line(1,0){2}}
\put(0,0){\line(0,1){2}}
\put(1,0){\line(0,1){2}}
\put(2,0){\line(0,1){2}}
\put(0.5,0.25){\makebox[0pt][c]{$1$}}
\put(0.5,1.25){\makebox[0pt][c]{$1$}}
\put(1.5,0.25){\makebox[0pt][c]{$1$}}
\put(1.5,1.25){\makebox[0pt][c]{$1$}}
\end{picture}
\quad
\cdots
\]
The connection between the definitions of recurrent by toppling
covering and by absence of FSC's
is given by the following statement, slightly more general that what
would suffice at this purpose.

\begin{prop}
\label{prop.oldP5}
Let $\vec{u} \in S'_+$ and $\vec{v} = t_{i_k} t_{i_{k-1}} \cdots
t_{i_1} \vec{u}$. Define 
$A = \bigcup_{1 \leq a \leq k} \{i_a\}$.
Then, for all $B$ such that $|B \setminus A| \leq 1$, 
$\vec{v}|_{B} \not\preceq \vec{f}_{\rm max}(B)$.
\end{prop}
This proposition implies as a corollary (for $|B\setminus A| = 0$)
that configurations which are recurrent by toppling covering are also
recurrent by absence of FSC's. The case $|B\setminus A| = 1$
also emerges naturally from the proof. 


\proof The claim $\vec{v}|_{B} \not\preceq \vec{f}_{\rm max}(B)$ can
be restated as the existence of $s \in B$ such that 
$v_s > f_{\rm max}(B)_s$. We will produce a valid choice for $s$.

If $|B\setminus A| = 1$, choose as $s$ the only site in $B$ and not in
$A$. Let $\vec{u}'=\vec{u}$ and $\tau(s)=0$ in this case.  Otherwise,
for all $i \in B$, call $\tau(i)$ the maximum $1 \leq a \leq k$ such
that $i_a = i$, then choose as $s$ the index realising the minimum of
$\tau(i)$, i.e., the site of $B$ that has performed its last toppling
more far in the past. Call $\vec{u}' = t_{i_{\tau(s)}} \cdots t_{i_1}
\vec{u}$, the configuration obtained after the last toppling in $s$.

Note that, as
$\vec{u} \in S'_+$ and this space is stable under topplings,
$u'_{s} \geq \ovl{z}_s-\Delta_{ss} + 1$.
In the remaining part of the avalanche, no more topplings occur at
$s$. Furthermore, all the other sites $j \in B$ do topple at least
once, at $a=\tau(j)$. Thus we have
\be
\begin{split}
v_s 
&= 
u'_s - 
\!\!\!
\sum_{a=\tau(s)+1}^k 
\!\!\!
\Delta_{i_a s}
\geq
\ovl{z}_s-\Delta_{ss} +1 - 
\!\!\!
\sum_{a=\tau(s)+1}^k 
\!\!\!
\Delta_{i_a s}
\\
&=
\ovl{z}_s- \sum_{i \in B} \Delta_{is}
+1
-
\!\!\!
\sum_{\substack{
\tau(s) < a \leq k
\\ a \not\in \{ \tau(j) \}_{j \in B}
}}
\!\!\!
\Delta_{i_a s}
\geq
\ovl{z}_s-
\!\!\!
\sum_{i \in B} 
\!\!\!
\Delta_{is} +1
\ef.
\end{split}
\ee
The comparison with the definition (\ref{eq.fmax}) of 
$\vec{f}_{\rm max}(B)$ allows to conclude. \qed

\myskip

We have now the ingredients to prove Theorem~\ref{theo:mainweak}.
 
\myskip

\noindent
{\sc Proof of Theorem \ref{theo:mainweak}.}
The two equations are related by the involution, so we only
prove (\ref{eq.intheo1w}). I.e., for all $\vec{z} \in S_+$, $a_i^\fdg
a^\dg_i a_i^\fdg \vec{z} = a_i^\fdg \vec{z}$.  First of all, as, for
$\vec{z} \in S_+$ and $\vec{w} \in \Omega$, $a_{\vec{w}} \vec{z} =
a_{\vec{w}} \relaxx(\vec{z})$, we can restrict our attention to
$\vec{z} \in S$. If $z_i < \ovl{z}_i$ we have $a^\dg_i a_i^\fdg
\vec{z} = \vec{z}$, and our relation follows. If $z_i = \ovl{z}_i$,
the avalanche due to the action of $a_i$ performs at least one
toppling at $i$. By Proposition~\ref{prop.oldP5}, $\vec{y} = a_i
\vec{z}$ has no FSC's with $I=\{i\}$ or $I=\{i,j\}$. That is, either
$y_i > \ovl{z}_i - \Delta_{ii}$, or $y_i = \ovl{z}_i - \Delta_{ii}$
and, for all $j \neq i$, $y_j \geq \ovl{z}_j - \Delta_{jj} -
\Delta_{ij}$. By direct inspection, in the first case 
$a_i^\fdg a^\dg_i \vec{y} = \hat{a}_i \hat{a}^{-1}_i \vec{y} =
\vec{y}$
and in the second case
$a_i^\fdg a^\dg_i \vec{y} = t^\fdg_i \hat{a}_i t^\dg_i \hat{a}^{-1}_i
\vec{y} = \vec{y}$.
In both cases, our relation follows.
\qed

\myskip
For a sandpile $\cA(\Delta,\ovl{\vec{z}},\udl{\vec{z}})$, call
$\cA|_I$ the sandpile described by the toppling matrix $\Delta_{I,I}$
the principal minor of $\Delta$ corresponding to the rows and columns
in $I$, and by the threshold vectors $\ovl{\vec{z}}|_I$ and
$\udl{\vec{z}}|_I$.  Also call $\vec{b}^+(I)$ the vector $\vec{b}^+$
associated to $\Delta_{I,I}$.  The observation that ultimately allows
to relate the definitions of recurrency by absence of FSC's and by
identity test is the fact that
\be
\label{eq.76545678}
\vec{f}_{\rm max}(I)
+
\vec{b}^+(I)
=
\ovl{\vec{z}}|_I
\ef.
\ee
A first remark in this direction is that, if $I=[n]$ is
not a FSC for $\vec{z}$, we can at least start the avalanche in the
definition of recurrent by identity test, i.e.\ we have at least one
site $i=\sigma(1)$ which is unstable. Indeed, we have at least one $i$
such that $z_i > f_{\rm max}([n])_i$. Then, by (\ref{eq.76545678}),
$(z+b^+)_i > f_{\rm max}([n])_i + b^+_i = \ovl{z}_i$.

As (\ref{eq.76545678}) holds for any set $I$, the reasoning above
works for any restricted sandpile $\cA|_I$.  Suppose to have a
configuration $\vec{z}^{(I)}$ which is recurrent in $\cA|_I$ by
absence of FSC's. Then we have at least one site $i\in I$ which is
unstable, because we have at least one $i$ such that $z^{(I)}_i >
f_{\rm max}(I)_i$, and $(z^{(I)}+b^+(I))_i > f_{\rm max}(I)_i +
b^+(I)_i = \ovl{z}_i$.

This allows to construct an induction. Let $I' = I
\setminus i$, $\vec{v}^{(I')} = (t_i \vec{z}^{(I)})|_{I'}$ and
$\vec{z}^{(I')} = \relaxx(\vec{v}^{(I')})$.  For $j \in I'$,
$v^{(I')}_j = z^{(I)}_j - \Delta_{ij} \geq z^{(I)}_j$.  
Remark that
the definition of $\vec{f}_{\rm max}(J)$ is the same on
any restricted sandpile $\cA_I$ with $I \supseteq J$.  Thus, as
$\vec{z}^{(I)}$ is recurrent for absence of FSC's,
and this property is preserved under relaxation (by Proposition
\ref{prop.oldP5}), also $\vec{z}^{(I')}$ is recurrent by absence of
FSC's, on $\cA|_{I'}$.  This, together with equation
(\ref{eq.76545678}), gives the induction step.

In summary, we can perform the complete avalanche $t_{\sigma(n)}
\cdots t_{\sigma(1)}$, algorithmically, by initialising $\vec{z}^{(0)}
= \vec{z} + \vec{b}^+$ and $I_0 = [n]$, and, for $a=0,\ldots, n-1$,
$i_a$ is any site $i$ at which $\vec{z}^{(a)}$ is unstable (which is
proven to exist by the reasoning above), $I_{a+1} = I_a \setminus
i_a$, and $\vec{z}^{(a+1)} = \relaxx( t_{i_a} \vec{z}^{(a)})$.

This completes the equivalence of our three definitions of recurrent
configurations, thus from now on we will omit to specify the defining
property.




\myskip
A directed graph can be associated to a sandpile, such that
$-\Delta_{ij}$ directed edges connect $i$ to $j$. This completely
encodes the off-diagonal part of $\Delta$. The remaining parameters,
in particular $\vec{b}^+$ and $\vec{b}^-$, can be encoded through
directed edges incoming from, our outgoing to, a special {\em sink} vertex.
In the
undirected case, $\Delta = \Delta^{\rm T}$, a bijection, called
\emph{Burning Test}, 
relates stable recurrent
configurations to spanning trees ~\cite{Dhar2}. A generalisation, called 
\emph{Script Algorithm}~\cite{Spe93}, extends this result also to directed graphs even 
in presence of greedy sites.
Through Kirchhoff Matrix-Tree Theorem, one thus gets in this
case that the number of stable recurrent configurations is $\det
\Delta$.

The configuration $\vec{p} = \ovl{\vec{z}} + \vec{1} -
\relaxx(\ovl{\vec{z}} + \vec{1})$ has two interesting properties: $p_i
\geq 1$ for all $i$, and $\vec{p} \sim \vec{0}$. This implies that,
for every $\vec{z} \in \mathbb{Z}^n$, the iteration of the map
$\vec{z} \to \relaxx(\vec{z}+\vec{p})$ must reach a fixed point, in
$R$, the subset of $S$ containing recurrent configurations (this would
be true also with $\vec{b}^+$ instead of $\vec{p}$, but slightly
harder to prove). Indeed, calling $c=\max_i (\ovl{z}_i - z_i + 1)$, as
$\relaxx(\relaxx(\cdots\relaxx(\vec{z}+\vec{p})\cdots
+\vec{p})+\vec{p}) = \relaxx(\vec{z}+c\,\vec{p})$, and
$\vec{z}+c\,\vec{p}$ is unstable at all sites,
$\relaxx(\vec{z}+c\,\vec{p})$ is both stable and recurrent (by
toppling covering), so it must be in $R$.

This reasoning proves that each equivalence class has at least one
representative in $R$. In the case $\vec{b}^+ \in \Omega$, as both the
cardinality of $R$ and the number of classes are $\det \Delta$, each
equivalence class must have a unique representative in
$R$.
In particular, the representative in $R$ of $\vec{0}$ is called the
\emph{recurrent identity}, and we denote it with the symbol $\id_r$.
This configuration can be found as the fixed point of the map $\vec{z}
\to \relaxx(\vec{z}+\vec{b}^+)$, started from $\vec{0}$~\cite{Creutz,Creutz2}, or of the map $\vec{z} \to
\relaxx(\vec{z}+\vec{p})$, or, under the mild assumption that
$\ovl{\vec{z}} \in \Omega$,
more directly, with no need of iterations, by the relation~\cite{LeBRossin}
\be
\id_r = \relaxx(\ovl{\vec{z}} + (\ovl{\vec{z}} - \relaxx(\ovl{\vec{z}} +
\ovl{\vec{z}})))
\ef.
\ee
Indeed, as
$\ovl{\vec{z}} - \relaxx(\ovl{\vec{z}} + \ovl{\vec{z}}) \in \Omega$, 
$\ovl{\vec{z}} + (\ovl{\vec{z}} - \relaxx(\ovl{\vec{z}} +
\ovl{\vec{z}}))$ is recurrent, thus its relaxation is in~$R$.

\myskip
In the $\udl{\vec{z}} = \vec{0}$ gauge,
the operation $\vec{u} \oplus \vec{v} := \relaxx(\vec{u} + \vec{v})$
sends $S \times S$ into $S$, and thus defines a semigroup on this
space. Furthermore, this operation also sends $S \times R \to R$, $R
\times S \to R$, and, as a corollary, $R \times R \to R$.

The charges behave linearly under this operation: $\vec{Q}(\vec{u}
\oplus \vec{v}) - \vec{Q}(\vec{u}) - \vec{Q}(\vec{v}) \in
\mathbb{Z}^n$.  The unicity of representatives in $R$ of the
equivalence classes allows then to construct inverses, for the action
on this space, and thus to promote $\oplus$ to a group action on $R$,
of which $\id_r$ is the group identity.  This structure was first
introduced by Creutz \cite{Creutz,Creutz2}, and then investigated in several
papers~\cite{Cori,Dhar_alg,LeBRossin,noi}.

The covariance of the notations allows to define the operation
$\oplus$ in general.
We have
\begin{align}
\vec{u} \oplus \vec{v} 
&:= 
\relaxx(\vec{u} + \vec{v} - \udl{\vec{z}})
\ef;
&
\vec{u} \oplus^\dg \vec{v} 
&:= 
\relaxx^\dg(\vec{u} + \vec{v} - \ovl{\vec{z}})
\ef.
\end{align}
Note however that now the charges behave in an affine way,
and only the translated versions, $\vec{Q}'(\vec{u}) =
\vec{Q}(\vec{u}) - \vec{Q}(\udl{\vec{z}})$ and $\vec{Q}''(\vec{u}) =
\vec{Q}(\vec{u}) - \vec{Q}(\ovl{\vec{z}})$ respectively for the two
operations, have no offset.  In particular, the two groups induced by
$\oplus$ and $\oplus^\dg$, as well as the two sets $R$ and $R^\dg$,
are isomorphic but not element-wise coincident, as the only natural
bijection among the two makes use of the involution $\iota$. 

\myskip
Call $\cM=\cM[a^\fdg_i,a^\dg_i]$ the transition monoid generated by
the $a_i$'s and $a^\dg_i$'s acting on our set of configurations. A
generic element in $\cM$ has the form
\be
A = 
a^\dg_{i_1^1} \cdots a^\dg_{i_{\ell(1)}^1} 
a^\fdg_{i_1^2} \cdots a^\fdg_{i_{\ell(2)}^2} 
a^\dg_{i_1^3} \cdots a^\dg_{i_{\ell(3)}^3} 
\;
\cdots
\;
a^\fdg_{i_1^{2k}} \cdots a^\fdg_{i_{\ell(2k)}^{2k}} 
\ef.
\ee
\section{Multitopplings in Abelian Sandpiles}
\label{sec:multitop}


Consider a situation not dissimilar from the BTW sandpile, with height
variables $z_i$ at the sites of a square lattice, but let the toppling
$t_i$ occur if the \emph{gradient} passes a given threshold. For
example, choose to perform a toppling at
$i$  if $i$ is a local maximum.

Despite the simplicity of such a rule, abelianity is lost, as is
easily verified. For example,  on a configuration
$\vec{z}$ as below, the two sites $i$ and $j$ are unstable, but $t_i
\vec{z}$ and $t_j \vec{z}$ differ, and it can well be that both have
no unstable sites.
\[
\setlength{\unitlength}{15pt}
\begin{picture}(4,4)
\put(1,0){\line(1,0){1}}
\put(0,1){\line(1,0){3}}
\put(0,2){\line(1,0){4}}
\put(1,3){\line(1,0){3}}
\put(2,4){\line(1,0){1}}
\put(0,1){\line(0,1){1}}
\put(1,0){\line(0,1){3}}
\put(2,0){\line(0,1){4}}
\put(3,1){\line(0,1){3}}
\put(4,2){\line(0,1){1}}
\put(1.5,0.3){\makebox[0pt][c]{6}}
\put(0.5,1.3){\makebox[0pt][c]{4}}
\put(1.5,1.3){\makebox[0pt][c]{5}}
\put(2.5,1.3){\makebox[0pt][c]{9}}
\put(1.5,2.3){\makebox[0pt][c]{8}}
\put(2.5,2.3){\makebox[0pt][c]{9}}
\put(3.5,2.3){\makebox[0pt][c]{6}}
\put(2.5,3.3){\makebox[0pt][c]{4}}
\put(1.1,1.15){\makebox[0pt][l]{\scriptsize{$i$}}}
\put(2.1,2.15){\makebox[0pt][l]{\scriptsize{$j$}}}
\end{picture}
\quad
\raisebox{7mm}{$\stackrel{\displaystyle{t_i}}{\longleftarrow}$}
\quad
\begin{picture}(4,4)
\put(1,0){\line(1,0){1}}
\put(0,1){\line(1,0){3}}
\put(0,2){\line(1,0){4}}
\put(1,3){\line(1,0){3}}
\put(2,4){\line(1,0){1}}
\put(0,1){\line(0,1){1}}
\put(1,0){\line(0,1){3}}
\put(2,0){\line(0,1){4}}
\put(3,1){\line(0,1){3}}
\put(4,2){\line(0,1){1}}
\put(1.5,0.3){\makebox[0pt][c]{5}}
\put(0.5,1.3){\makebox[0pt][c]{3}}
\put(1.5,1.3){\makebox[0pt][c]{9}}
\put(2.5,1.3){\makebox[0pt][c]{8}}
\put(1.5,2.3){\makebox[0pt][c]{7}}
\put(2.5,2.3){\makebox[0pt][c]{9}}
\put(3.5,2.3){\makebox[0pt][c]{6}}
\put(2.5,3.3){\makebox[0pt][c]{4}}
\put(1.1,1.15){\makebox[0pt][l]{\scriptsize{$i$}}}
\put(2.1,2.15){\makebox[0pt][l]{\scriptsize{$j$}}}
\end{picture}
\quad
\raisebox{7mm}{$\stackrel{\displaystyle{t_j}}{\longrightarrow}$}
\quad
\begin{picture}(4,4)
\put(1,0){\line(1,0){1}}
\put(0,1){\line(1,0){3}}
\put(0,2){\line(1,0){4}}
\put(1,3){\line(1,0){3}}
\put(2,4){\line(1,0){1}}
\put(0,1){\line(0,1){1}}
\put(1,0){\line(0,1){3}}
\put(2,0){\line(0,1){4}}
\put(3,1){\line(0,1){3}}
\put(4,2){\line(0,1){1}}
\put(1.5,0.3){\makebox[0pt][c]{5}}
\put(0.5,1.3){\makebox[0pt][c]{3}}
\put(1.5,1.3){\makebox[0pt][c]{9}}
\put(2.5,1.3){\makebox[0pt][c]{9}}
\put(1.5,2.3){\makebox[0pt][c]{8}}
\put(2.5,2.3){\makebox[0pt][c]{5}}
\put(3.5,2.3){\makebox[0pt][c]{7}}
\put(2.5,3.3){\makebox[0pt][c]{5}}
\put(1.1,1.15){\makebox[0pt][l]{\scriptsize{$i$}}}
\put(2.1,2.15){\makebox[0pt][l]{\scriptsize{$j$}}}
\end{picture}
\]
We describe now another variant of the BTW sandpile in which
abelianity is lost. We come back to toppling criteria based on having
heights above given thresholds, as in the ordinary BTW (and not
gradients, as above). Now the rule is that, if two neighbouring sites
have both height 3 or higher, both their heights are decreased by 3, and
the height of each of the 6 neighbours is increased by 1. Again, it is
not difficult to imagine situations in which two sites $i$ and $j$ are
unstable, but $t_i \vec{z}$ and $t_j \vec{z}$ differ, and both have no
unstable sites, as in the configuration below.
\[
\setlength{\unitlength}{15pt}
\begin{picture}(4,4)
\put(1,0){\line(1,0){2}}
\put(0,1){\line(1,0){4}}
\put(0,2){\line(1,0){4}}
\put(1,3){\line(1,0){3}}
\put(2,4){\line(1,0){1}}
\put(0,1){\line(0,1){1}}
\put(1,0){\line(0,1){3}}
\put(2,0){\line(0,1){4}}
\put(3,0){\line(0,1){4}}
\put(4,1){\line(0,1){2}}
\put(1.5,0.3){\makebox[0pt][c]{3}}
\put(2.5,0.3){\makebox[0pt][c]{2}}
\put(0.5,1.3){\makebox[0pt][c]{1}}
\put(1.5,1.3){\makebox[0pt][c]{1}}
\put(2.5,1.3){\makebox[0pt][c]{0}}
\put(3.5,1.3){\makebox[0pt][c]{2}}
\put(1.5,2.3){\makebox[0pt][c]{2}}
\put(2.5,2.3){\makebox[0pt][c]{4}}
\put(3.5,2.3){\makebox[0pt][c]{2}}
\put(2.5,3.3){\makebox[0pt][c]{1}}
\put(1.1,1.15){\makebox[0pt][l]{\scriptsize{$i$}}}
\put(2.1,1.15){\makebox[0pt][l]{\scriptsize{$k$}}}
\put(2.1,2.15){\makebox[0pt][l]{\scriptsize{$j$}}}
\end{picture}
\quad
\raisebox{7mm}{$\stackrel{\displaystyle{t_{ik}}}{\longleftarrow}$}
\quad
\begin{picture}(4,4)
\put(1,0){\line(1,0){2}}
\put(0,1){\line(1,0){4}}
\put(0,2){\line(1,0){4}}
\put(1,3){\line(1,0){3}}
\put(2,4){\line(1,0){1}}
\put(0,1){\line(0,1){1}}
\put(1,0){\line(0,1){3}}
\put(2,0){\line(0,1){4}}
\put(3,0){\line(0,1){4}}
\put(4,1){\line(0,1){2}}
\put(1.5,0.3){\makebox[0pt][c]{2}}
\put(2.5,0.3){\makebox[0pt][c]{1}}
\put(0.5,1.3){\makebox[0pt][c]{0}}
\put(1.5,1.3){\makebox[0pt][c]{4}}
\put(2.5,1.3){\makebox[0pt][c]{3}}
\put(3.5,1.3){\makebox[0pt][c]{1}}
\put(1.5,2.3){\makebox[0pt][c]{1}}
\put(2.5,2.3){\makebox[0pt][c]{3}}
\put(3.5,2.3){\makebox[0pt][c]{2}}
\put(2.5,3.3){\makebox[0pt][c]{1}}
\put(1.1,1.15){\makebox[0pt][l]{\scriptsize{$i$}}}
\put(2.1,1.15){\makebox[0pt][l]{\scriptsize{$k$}}}
\put(2.1,2.15){\makebox[0pt][l]{\scriptsize{$j$}}}
\end{picture}
\quad
\raisebox{7mm}{$\stackrel{\displaystyle{t_{jk}}}{\longrightarrow}$}
\quad
\begin{picture}(4,4)
\put(1,0){\line(1,0){2}}
\put(0,1){\line(1,0){4}}
\put(0,2){\line(1,0){4}}
\put(1,3){\line(1,0){3}}
\put(2,4){\line(1,0){1}}
\put(0,1){\line(0,1){1}}
\put(1,0){\line(0,1){3}}
\put(2,0){\line(0,1){4}}
\put(3,0){\line(0,1){4}}
\put(4,1){\line(0,1){2}}
\put(1.5,0.3){\makebox[0pt][c]{2}}
\put(2.5,0.3){\makebox[0pt][c]{2}}
\put(0.5,1.3){\makebox[0pt][c]{0}}
\put(1.5,1.3){\makebox[0pt][c]{5}}
\put(2.5,1.3){\makebox[0pt][c]{0}}
\put(3.5,1.3){\makebox[0pt][c]{2}}
\put(1.5,2.3){\makebox[0pt][c]{2}}
\put(2.5,2.3){\makebox[0pt][c]{0}}
\put(3.5,2.3){\makebox[0pt][c]{3}}
\put(2.5,3.3){\makebox[0pt][c]{2}}
\put(1.1,1.15){\makebox[0pt][l]{\scriptsize{$i$}}}
\put(2.1,1.15){\makebox[0pt][l]{\scriptsize{$k$}}}
\put(2.1,2.15){\makebox[0pt][l]{\scriptsize{$j$}}}
\end{picture}
\]
In this section we will describe a family of sandpiles for which
abelianity is preserved.  In these models we have toppling rules based
on ``heights above given thresholds'', just as in the ordinary ASM and
in the example above. In fact, these models generalise the ordinary
ASM, because we have simultaneously all the ordinary toppling moves,
and a (possibly empty) collection of \emph{multitoppling} moves
satisfying certain criteria.

As we will see later on, the moves described in the second example
would be a valid choice -- and remark indeed how, including also
ordinary topplings, the configurations on the left and right of the
figure above are still unstable on $j$ and $i$, respectively, and
become identical after these topplings.

Consider an ordinary ASM $\cA(\Delta,\ovl{\vec{z}},\udl{\vec{z}})$.
For simplicity of notations, we set in the positive-cone gauge, so
that site $i$ is unstable if $z_i \geq \Delta_{ii}$.

For a non-empty set $I\subseteq [n]$, call 
$\vec{\Delta}_I = \sum_{i \in I} \vec{\Delta}_i$.  A multitoppling
operator $t_I$ can be associated to a set $I$. First of all, $\vec{z}$
is unstable for toppling $I$ if 
$z_j \geq (\Delta_I)_j$ for all $j\in I$ 
(note how, in the positive-cone gauge, this coincides with the
ordinary definition when $|I|=1$). Then, if the configuration is
unstable, it is legitimate to perform the toppling $t_I \vec{z} =
\vec{z} - \vec{\Delta}_I$. Also note that the absence of greedy 
sites implies that, for all $j \in I$, $(\Delta_I)_j \geq 0$,
while for all $j \not\in I$, $(\Delta_I)_j \leq 0$.

Consider a collection $\cL$ of non-empty subsets of $[n]$.
The interest in multitoppling rules for the Abelian Sandpile
Model is in the following fact
\begin{prop}
\label{prop.diamondMulti}
Suppose that, for every $I, J \in \cL$, the sets $I' = I \setminus J$
and $J' = J \setminus I$ are either empty or in $\cL$. Then, if
$\vec{z}$ is unstable for both $I$ and $J$, $t_I \vec{z}$ is unstable
for $J'$ and $t_J \vec{z}$ is unstable for $I'$.
\end{prop}
As clearly $t_{J'} t_I \vec{z}=t_{I'} t_J \vec{z}$, it easily follows
\begin{coroll}
In the conditions of Proposition \ref{prop.diamondMulti}, the operator
$\relaxx$ is unambiguous.
\end{coroll}
Antitopplings $t^\dg_I$ and antirelaxation $\relaxx^\dg$ are defined
just as in the ordinary case, e.g.\ through the involution $\iota$,
which is still defined as in (\ref{eq.iota}). A configuration is
stable if no toppling or antitoppling can occur (this coincides with
the definition of $S$ in the ordinary case). Note that, if $\cL$ does
not contain the atomic set $\{i\}$, $S$ is either empty or of infinite
cardinality (because, if $\vec{z} \in S$, also $\vec{z}+c\; \vec{e}_i
\in S$ for any $c \in \mathbb{Z}$).  In order to exclude this
pathological case, we will assume in the following that $\cL$ includes
the set $\cL_0 = \big\{ \{i\} \big\}_{i \in [n]}$ of all atomic
subsets, i.e.\ single-site topplings. In this case, as stability
w.r.t.\ $\cL$ is a more severe requirement than stability
w.r.t.\ $\cL' \subset \cL$, we have that $S$ is a subset of the set
$S_0$ of stable configurations in the associated sandpile with only
single-site topplings, and thus of finite cardinality. As we will see
later on, for any set $\cL$ as in Proposition \ref{prop.diamondMulti}
the set $S$ is non-empty, and actually contains a set isomorphic to
$R$, thus it has cardinality bounded below by $|R_0| = \det \Delta$,
an above by $|S_0| = \prod_i (\ovl{z}_i - \udl{z}_i +1)$.

If we require both that $\cL \supseteq \cL_0$, and satisfies the
hypotheses of Proposition \ref{prop.diamondMulti}, we get that $\cL$
is a down set in the lattice of subsets, that is, for all $I \in \cL$
and $H \subseteq \cL$ non-empty, also $H \in \cL$ (this is trivially
seen: with notations as in Proposition \ref{prop.diamondMulti}, take
$I$ and $J=I \setminus H$).

For a multitoppling sandpile $\cA=\cA(\cL)$, we will call $\cA_0 =
\cA(\cL_0)$ the associated single-site toppling sandpile.

The concept of recurrent configuration has to be reanalysed in this
context. The various alternate definitions are modified (in the more
complicated situation, but under the simplification of the choice of
positive-cone gauge) into

\begin{defin}
A configuration $\vec{z}$ is \emph{recurrent by identity test} if
there exists an ordered sequence $(I_1, \ldots, I_k)$ of subsets of
$[n]$, constituting a partition of $[n]$ (i.e.\ for all $i \in [n]$
there exists a unique $a$
such that $i \in I_a$), such
that $t_{I_k} \cdots t_{I_2} t_{I_1} (\vec{z} + \vec{b}^+) = \vec{z}$
is a valid toppling sequence.
\end{defin}

\begin{defin}
A configuration $\vec{z}$ is \emph{recurrent by toppling covering} if
there exists a configuration $\vec{u}$,
such that
$\vec{z} = t_{I_k} \cdots t_{I_1} \vec{u}$ is a valid toppling
sequence, and each site $i$ is contained in at least one of the
$I_a$'s.
\end{defin}

\begin{defin}
A configuration $\vec{z}$ is \emph{recurrent by absence of FSC's} if,
for every set $I$ of sites, there exists $J \in \cL$ with $L = I \cap J
\neq \emptyset$, and $z_j \geq \sum_{i \in I \setminus L} 
\Delta_{ij}$ for all $j \in L$.
\end{defin}

\noindent
All the reasonings are the immediate generalisation of the ones
already done in Section \ref{ssec:further}. We just report here the
appropriate modifications in Proposition \ref{prop.oldP5}
(recall that in our gauge $\Omega \equiv S'_+$, and is left stable by
the topplings).

\begin{prop}
\label{prop.oldP5MT}
Let $\vec{u} \in \Omega$ and $\vec{v} = t_{I_k} t_{I_{k-1}} \cdots
t_{I_1} \vec{u}$. Define $A = \bigcup_{1 \leq a \leq k} I_a$.
For any set $B$, there exists a non-empty set $H \subseteq B$, such
that $B \subseteq H \cup A$, and, for all $s \in H$,
$v_s > -\sum_{i \in (A \cap B) \setminus H} \Delta_{is}$.
\end{prop}



\proof We will produce explicitly a valid choice of $H$.
If $B \nsupseteq A$, let $H = B \setminus A$ and $\vec{u}'=\vec{u}$.  
Note that $(B \cap A) \setminus H = B \cap A$ in this case.
If $B \supseteq A$, for all $i \in B$, call $\tau(i)$ the maximum $1
\leq a \leq k$ such that $i \in I_a$, then call $\tau = \max_i
\tau(i)$, and $J = I_{\tau(i)}$, i.e., the multitoppling that covered
any portion of $B$ more far in the past. Call $\vec{u}' = t_{I_{\tau}}
\cdots t_{I_1} \vec{u}$, the configuration obtained after this last
multitoppling.  
Note that all the entries of $\vec{u}'$ are non-negative.
In the remaining part of the avalanche, for some
non-empty set $H \subseteq J$, no more topplings occur.  Conversely,
all the sites $j \in (A\cap B) \setminus H$ do topple at least once
(possibly in a multitoppling event). Thus we have, for each $s \in
H$,
\be
\begin{split}
v_s
&\geq
- \sum_{i \in (B \cap A) \setminus H}
\Delta_{is}
\ef,
\end{split}
\ee
as was to be proven.  \qed

We should modify the concept of forbidden sub-configuration along the
same lines. We want to produce pairs $(I,\vec{f})$ such that,
if $\vec{z}$ is recurrent by toppling covering, $\vec{z}|_I \neq \vec{f}$.
For a given $I$, a vector $\vec{f}$ has the property above if, for all
$J \in \cL$, $J \subseteq I$,
$\vec{f}|_J \preceq -\sum_{k \in I \setminus J} \vec{\Delta}_{k}|_J$.
Note that the fact that $\cL$ is a down set has been used to restrict
the set of $J$'s to analyse.

A collection of the forbidden pairs $(I,\vec{f})$ with smallest $|I|$,
such that no other $\vec{f}'$ exists with $\vec{f}' \succ \vec{f}$ and
$(I,\vec{f}')$ a forbidden pair, in the BTW sandpile with
multitoppling rules on all pairs of adjacent sites, is as follows:
\[
\setlength{\unitlength}{12.5pt}
\begin{picture}(1,1)
\put(0,0){\line(1,0){1}}
\put(0,1){\line(1,0){1}}
\put(0,0){\line(0,1){1}}
\put(1,0){\line(0,1){1}}
\put(0.5,0.3){\makebox[0pt][c]{\scriptsize{$-1$}}}
\end{picture}
\qquad
\begin{picture}(3,1)
\put(0,0){\line(1,0){3}}
\put(0,1){\line(1,0){3}}
\put(0,0){\line(0,1){1}}
\put(1,0){\line(0,1){1}}
\put(2,0){\line(0,1){1}}
\put(3,0){\line(0,1){1}}
\put(0.5,0.25){\makebox[0pt][c]{$0$}}
\put(1.5,0.25){\makebox[0pt][c]{$0$}}
\put(2.5,0.25){\makebox[0pt][c]{$0$}}
\end{picture}
\qquad
\begin{picture}(2,2)
\put(0,0){\line(1,0){2}}
\put(0,1){\line(1,0){2}}
\put(0,2){\line(1,0){1}}
\put(0,0){\line(0,1){2}}
\put(1,0){\line(0,1){2}}
\put(2,0){\line(0,1){1}}
\put(0.5,0.25){\makebox[0pt][c]{$0$}}
\put(0.5,1.25){\makebox[0pt][c]{$0$}}
\put(1.5,0.25){\makebox[0pt][c]{$0$}}
\end{picture}
\quad
\cdots
\quad
\begin{picture}(2,2)
\put(0,0){\line(1,0){2}}
\put(0,1){\line(1,0){2}}
\put(0,2){\line(1,0){2}}
\put(0,0){\line(0,1){2}}
\put(1,0){\line(0,1){2}}
\put(2,0){\line(0,1){2}}
\put(0.5,0.25){\makebox[0pt][c]{$1$}}
\put(0.5,1.25){\makebox[0pt][c]{$0$}}
\put(1.5,0.25){\makebox[0pt][c]{$0$}}
\put(1.5,1.25){\makebox[0pt][c]{$1$}}
\end{picture}
\qquad
\begin{picture}(2,2)
\put(0,0){\line(1,0){2}}
\put(0,1){\line(1,0){2}}
\put(0,2){\line(1,0){2}}
\put(0,0){\line(0,1){2}}
\put(1,0){\line(0,1){2}}
\put(2,0){\line(0,1){2}}
\put(0.5,0.25){\makebox[0pt][c]{$0$}}
\put(0.5,1.25){\makebox[0pt][c]{$1$}}
\put(1.5,0.25){\makebox[0pt][c]{$1$}}
\put(1.5,1.25){\makebox[0pt][c]{$0$}}
\end{picture}
\quad
\cdots
\]
Note that, at difference with the single-site sandpile, in general
there is not  a unique 
$\vec{f}_{\rm max}(I)$ such that $(I,\vec{f})$ is a forbidden pair if
and only if $\vec{f}\preceq \vec{f}_{\rm max}(I)$.

At the level of the monoid $\cM[a^\fdg_i,a^\dg_i]$, we have some extra
relations associated to multitoppling rules. For example, consider
the action on $S'_+$.
While in the single-site
sandpile we have relations (\ref{eq.ADeltaRela}), in the
multitoppling case we also have relations of the form, for each $I
\in \cL$,
\be
\label{eq.ADeltaRelaMT}
\prod_{j \in I} a_j^{(\Delta_{I})_j}
=
\prod_{j \not\in I} a_j^{-(\Delta_{I})_j}
\ef.
\ee
Note that multiplying left-hand and right-hand sides of relations
(\ref{eq.ADeltaRela}) for all $j \in I$ we would have got a weaker
relation, in which the two sides of (\ref{eq.ADeltaRelaMT}) are
multiplied by
\be
\prod_{\substack{
i \in I \\ j \in I \setminus i}}
a_j^{-\Delta_{ij}}
\ef,
\ee
(recall that at the level of the monoid it is not legitimate to take
inverses of $a_i$'s, and that, in the equations above, all the
exponents are indeed non-negative).

At the level of the equivalence relation $\sim$, and thus of charges
$\vec{Q}(\vec{z})$, nothing changes. In particular, it is easy to see
that $\vec{u} \sim \vec{v}$ in $\cA$ if and only if $\vec{u} \sim
\vec{v}$ in $\cA_0$.  This also leads to the fact that there is
exactly one stable recurrent configuration per equivalence class, and
that $\oplus$ defines a group structure over $R$, just as in the
single-site toppling sandpile $\cA_0 = \cA(\cL_0)$ associated to the
multitoppling sandpile $\cA=\cA(\cL)$.

Note however
that the set $R$ is \emph{different} from the set $R_0$ of stable
recurrent configurations for $\cA_0$, and the fact that they have the
same cardinality results from subtle compensations between
stable/unstable and recurrent/transient configurations. For example,
in the BTW sandpile with multitoppling rules on all pairs of adjacent
sites, the set $R$ loses the configurations with adjacent pairs
\hbox{\setlength{\unitlength}{10pt}
\begin{picture}(2.1,1)(0,0.15)
\put(0,0){\line(1,0){2}}
\put(0,1){\line(1,0){2}}
\put(0,0){\line(0,1){1}}
\put(1,0){\line(0,1){1}}
\put(2,0){\line(0,1){1}}
\put(0.5,0.25){\makebox[0pt][c]{\scriptsize{$3$}}}
\put(1.5,0.25){\makebox[0pt][c]{\scriptsize{$3$}}}
\end{picture}},
which are now unstable, but gains 
the configurations with adjacent pairs
\hbox{\setlength{\unitlength}{10pt}
\begin{picture}(2.1,1)(0,0.15)
\put(0,0){\line(1,0){2}}
\put(0,1){\line(1,0){2}}
\put(0,0){\line(0,1){1}}
\put(1,0){\line(0,1){1}}
\put(2,0){\line(0,1){1}}
\put(0.5,0.25){\makebox[0pt][c]{\scriptsize{$0$}}}
\put(1.5,0.25){\makebox[0pt][c]{\scriptsize{$0$}}}
\end{picture}},
which are not FSC's anymore.

A natural bijection between $R$ and $R_0$, preserving the group
structure, is obtained by associating to $\vec{z} \in R_0$ the
configuration $\relaxx(\vec{z}) \in R$, where $\relaxx$ is the
complete (multitoppling) relaxation.

In the case of a tight sandpile, the set $R^\dg$, containing the
conjugate of the stable recurrent configurations in the
single-toppling sandpile $\cA_0$, coincides with the set of recurrent
configurations of the sandpile in which $\cL$ contains all the subsets
of $[n]$. In this case, there are no stable transient configurations,
and the condition for $\vec{z}$ being stable w.r.t.\ any toppling $I$,
i.e.\ $\vec{z}|_I \not\succeq \vec{\Delta}_I$ for all non-empty $I
\subseteq [n]$, is related, by conjugation, to the condition of not
having FSC's in $\cA_0$, i.e., $(\iota\, \vec{z})|_I \not\preceq
\vec{f}_{\rm max}(I)$ (because $\vec{f}_{\rm max}(I) = (\ovl{\vec{z}}
- \vec{\Delta}_I)|_I$, and, in applying the definition of $\iota$, we
should recall that the multitoppling sandpile is formulated in the
positive-cone gauge).

\myskip
Now consider an {\em anomalous} relaxation process, $\rho_I$, which may
perform a multitoppling rule $I$ only at the first step, if possible,
and then perform a single-toppling relaxation with $\relaxx_0$. Such a
process is unambiguous, but different processes may not commute,
i.e.\ $\rho_I(\rho_J(\vec{z})) \neq \rho_J(\rho_I(\vec{z}))$ in
general.

Nonetheless, take a whatever semi-infinite sequence
$(I_1,I_2,I_3,\ldots)$ of elements in $\cL \setminus \cL_0$, such that
all elements in $\cL \setminus \cL_0$, occur infinitely-many times.
It is easy to see that, for all $\vec{z}$, there exists a truncation
time $t=t(\vec{z})$ such that, for all $s \geq t$, $\rho_{I_s}
\rho_{I_{s-1}} \cdots \rho_{I_1}(\vec{z}) = \relaxx(\vec{z})$, and in
particular $\rho_I \relaxx(\vec{z}) = \relaxx(\vec{z})$ for all $I \in
\cL$.

The interest in these processes $\rho_I$ is in the fact that their
action is strongly related to the action of the idempotents $a^\dg_i
a^\fdg_i$. Before stating and proving this relation in precise terms,
it is instructive to investigate first how this works in the case of
the BTW model. One easily recognizes that $a^\dg_i a^\fdg_i \vec{z} =
\vec{z}$ if $z_i < 3$, or $z_i = 3$ and $z_j < 3$ for all the
neighbours $j$ of $i$. In the first case, no topplings or
antitopplings are involved, while in the second case exactly one
toppling and one antitoppling at $i$ occur. Conversely, if $z_i = 3$
and $z_j = 3$ for some neighbour $j$, $a_i$ causes an avalanche for
which a valid sequence of topplings may start with $(i,j,\ldots)$,
i.e.\ $a_i \vec{z} = t_{i_k} \cdots t_{i_3} t_j t_i \vec{z}$ for some
$(i_3, \ldots, i_k)$. The \emph{effect} of the two initial topplings
is identical to the effect of a multitoppling at a pair
$\{i,j\}$. And, crucially, also the \emph{if condition} coincides with
the one for a configuration to be unstable w.r.t.\ the multitoppling
at $\{i,j\}$.  Thus, a configuration $\vec{z} \in \mathbb{Z}^n$ is
left stable by the application of $a^\dg_i a^\fdg_i$ if and only if it
is stable w.r.t.\ both the toppling $i$ and the multitopplings
$\{i,j\}$ for all $j$ neighbours of $i$. This proves 
Theorem \ref{theo.orchidweak} in the case of the BTW sandpile, and
characterizes $\vec{y}(\vec{z},I)$ as the result of the relaxation of
$\vec{z}$, in the multitoppling sandpile for which $\cL \setminus
\cL_0 = \big\{ \{i,j\} \big\}_{\textrm{$i \in I$, $d(i,j)=1$}}$\;.

The proof in the general setting, that we present below, is completely
analogous.

\noindent
{\sc Proof of Theorem \ref{theo.orchidweak}.}
One finds
that $a^\dg_i a^\fdg_i \vec{z} = \vec{z}$ if $z_i < \ovl{z}_i$, or
$z_i = \ovl{z}_i$ and $z_j \leq \ovl{z}_j + \Delta_{ij}$ for all $j
\neq i$. Again, in the first case no topplings or antitopplings are
involved, while in the second case exactly one toppling and one
antitoppling at $i$ occur.

Conversely, if the conditions above are violated, $a_i$ causes an
avalanche for which a valid sequence of topplings may start with
$(i,j,\ldots)$, and the effect of the two initial topplings is
identical to the effect of a multitoppling at a pair $\{i,j\}$.

The if condition for the multitoppling $\{i,j\}$ to occur is that $z_i
> \ovl{z}_i + \Delta_{ji}$ and $z_j > \ovl{z}_j + \Delta_{ij}$.  The
condition for the avalanche to involve topplings at $i$ and $j$ is
$z_i = \ovl{z}_i$ and $z_j > \ovl{z}_j + \Delta_{ij}$. These two sets
of conditions are certainly simultaneously not satisfied if
$\Delta_{ij} = 0$, thus we can restrict our attention to sites $j$
such that $\Delta_{ij} < 0$. In this case, the two sets coincide if
and only if $\Delta_{ji} = -1$. Thus, in order to make the sets
coincide for all sites $i$, we need that $\Delta$ is symmetric, and
all the non-zero off-diagonal entries are $-1$, as required in the
theorem hypotheses.

The configuration $\vec{y}(\vec{z},I)$ is the result of the relaxation of
$\vec{z}$, in the multitoppling sandpile for which $\cL \setminus
\cL_0 = \big\{ \{i,j\} \big\}_{i \in I, \Delta_{ij}=-1}$\,.
\qed

\section{Discussion of Markov Chain dynamics involving both sand addition
  and removal}
\label{sec:dynams}

In this section we discuss several dynamics involving the operators
$a^\fdg_i$ and $a^\dg_i$. Each example has a different theoretical and
phenomenological motivation, and is intended to describe a different
feature of out-of-equilibrium steady states. To keep the
visualisation simple, all our examples are variations of the BTW
model, on portions of the square lattice and with heights in the range
$\{0,1,2,3\}$.

\subsection{A reminder of the Karmakar and Manna protocol}
\label{ssec:dynKM}

The first and most natural dynamics in this family is the modification
of the BTW sandpile, where, with probability $p$ and
$1-p$  an $a^\fdg_i$ or an $a^\dg_i$ move is performed, at a randomly
chosen site $i$. This dynamics has been investigated in detail by
Karmakar and Manna in \cite{MannaPH}. For $p>1/2$ and $p<1/2$, the
system has features resembling the ones of the ordinary BTW model, and
of its symmetric image under the involution $z_i \leftrightarrow
3-z_i$. For example, for $p>1/2$ the avalanches have an algebraic-tail
distribution (the exponent varies with $p$), while the anti-avalanches
have a distribution with most of the support on values of order 1. Of
course, these two features are swapped for $p<1/2$.  In a window near
$p=p_c=1/2$, for which several scaling exponents have been investigated
numerically, aspects of a new criticality phenomenon emerge.
For example, at $p=p_c$ the
distribution of sizes of both avalanches and anti-avalanches seem to
follow a stretched-exponential.

As another example of new emerging features, both the variance and the
correlation times of the total mass in the system increase when $p_c$ is
approached, and the variation in $p$ of the average total mass
diverges at $p_c$, with a new scaling exponent.

\subsection{Dynamics preserving the total mass}
\label{ssec:dyntorus}

Our first dynamics is a minor modification of the one described above,
in the microcanonical ensemble with fixed total mass, at its critical
value. Among other things, this allows to use periodic boundary
conditions. Indeed, the dynamics studied by Karmakar and Manna, even
at $p=p_c$, if implemented on a graph with no boundary, may produce an
infinite avalanche or anti-avalanche, 
while in a dynamics with conserved mass, with randomized initialisation, if
the initial density is in the range $3-\rho_* < \rho < \rho_*$, where
$\rho_*$ is the critical density (having value $\rho^*=2.125288\ldots$
\cite{WilsonFLprl,WilsonFL}, note that, for a long time, it was mistakenly
believed that $\rho^*=17/8$), in the large volume limit we are protected
from infinite avalanches.

We want to emphasise here the behaviour of thermalisation process at short times,
that shows coarsening, and an aggregation phenomenon, which is purely
dynamical.

More generally, under some aspects, our dynamics compares to the one
of Karmakar and Manna roughly as Ising at fixed magnetisation under
Kawasaki dynamics relates to Ising heat-bath evolution. Recall however
that the presence of conserved quantities has more dramatic effects in
non-equilibrium systems than it has in Boltzmann equilibrium theory
(see e.g.\ the very different features of the Driven Lattice Gas, at
conserved number of particles or with open boundaries~\cite{ziabook}).

\begin{figure}
\[
\setlength{\unitlength}{89pt}
\begin{picture}(1.964,3)(0,3.05)
\put(-1,5.05){\includegraphics[scale=.33]{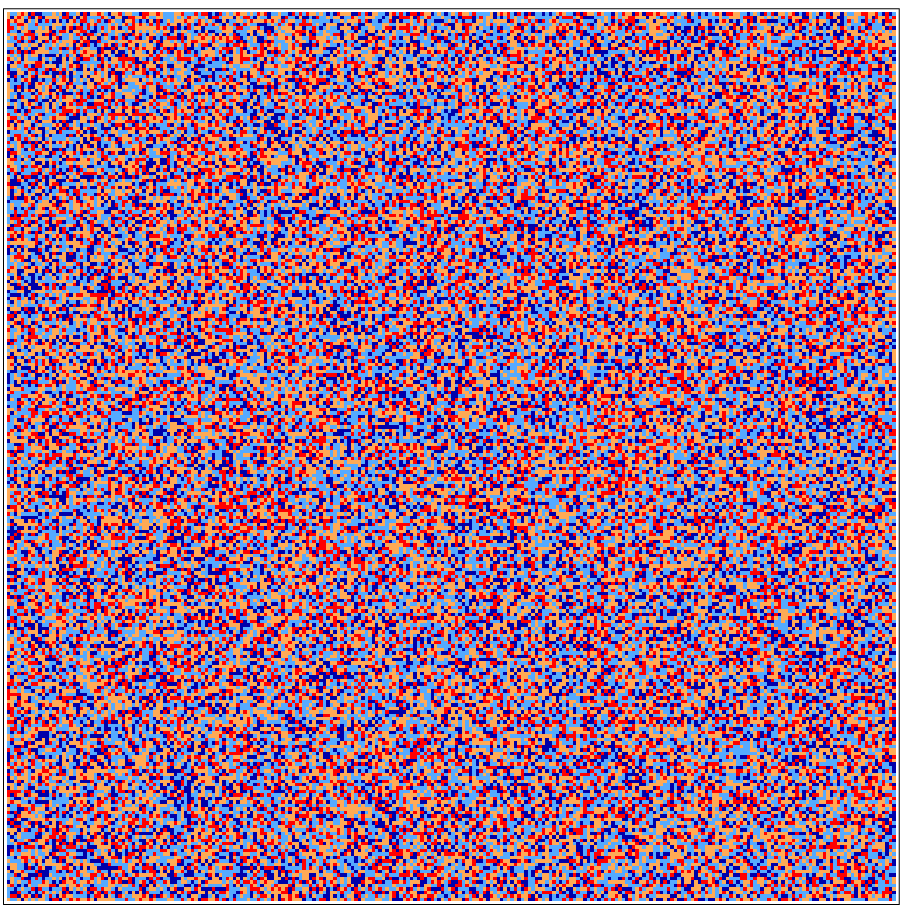}}
\put(0,5.05){\includegraphics[scale=.33]{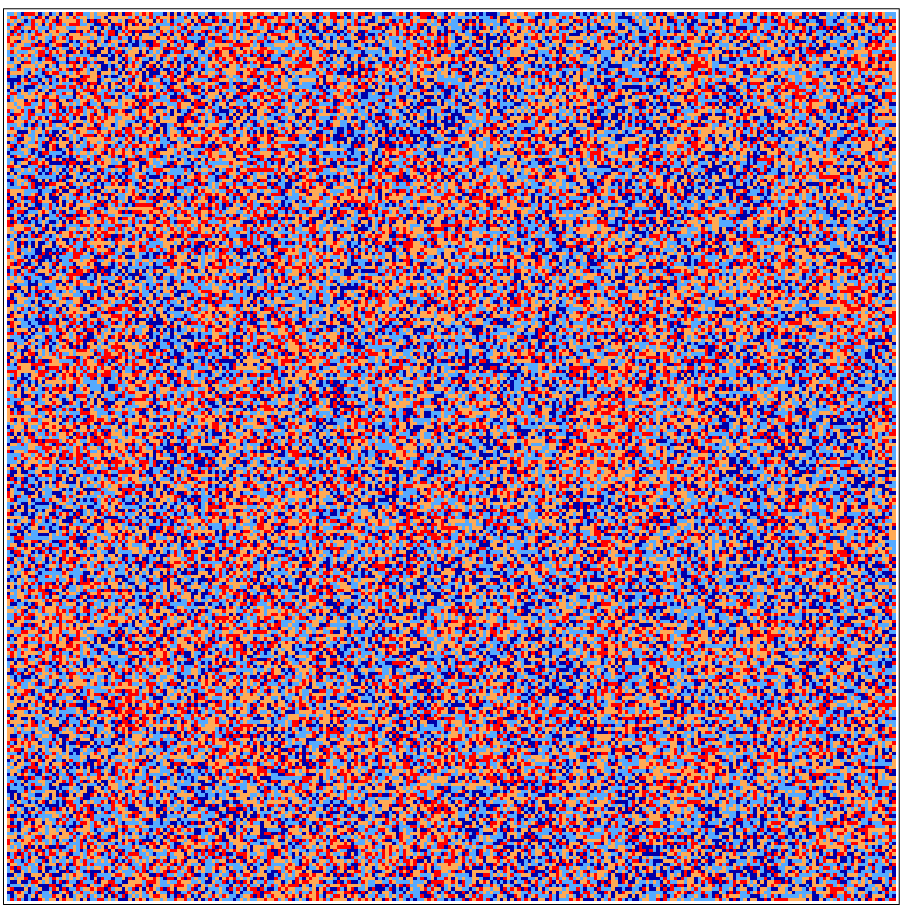}}
\put(1,5.05){\includegraphics[scale=.33]{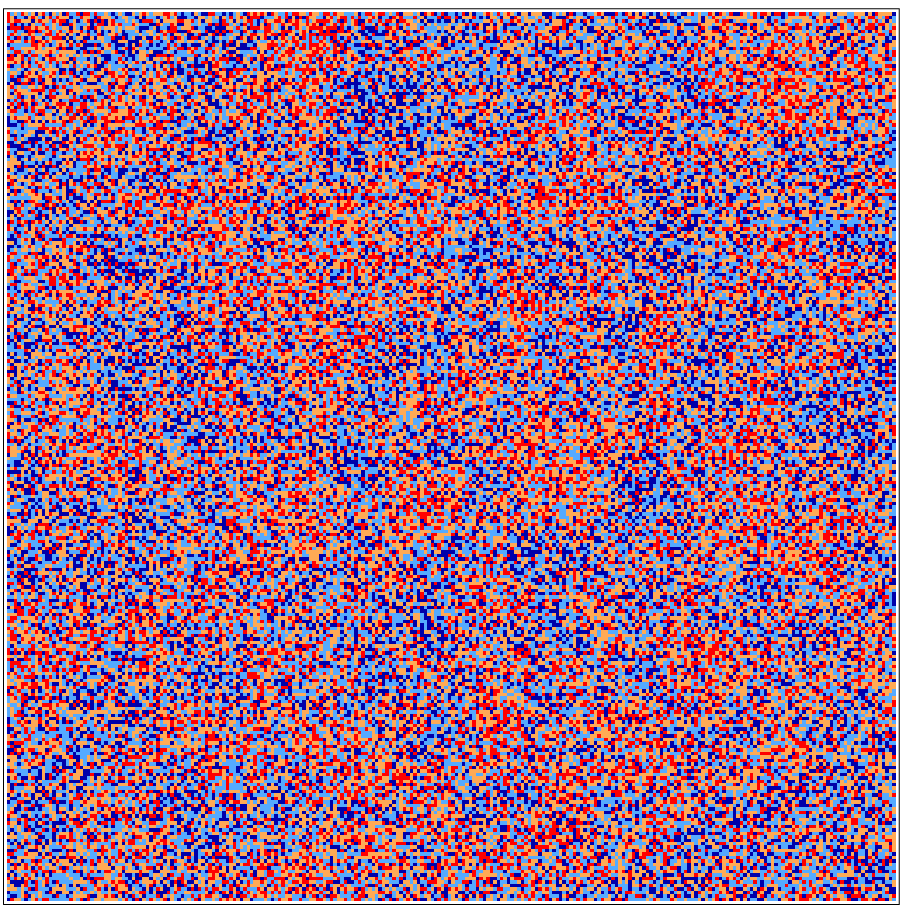}}
\put(2,5.05){\includegraphics[scale=.33]{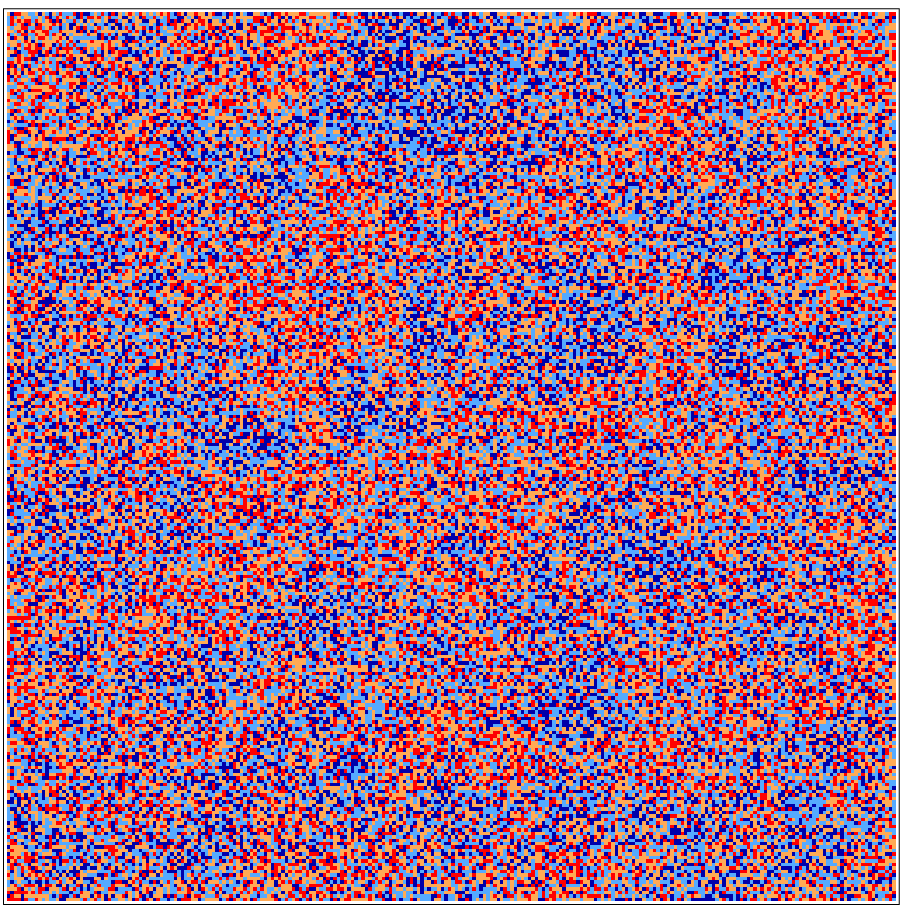}}

\put(-1,4.05){\includegraphics[scale=.33]{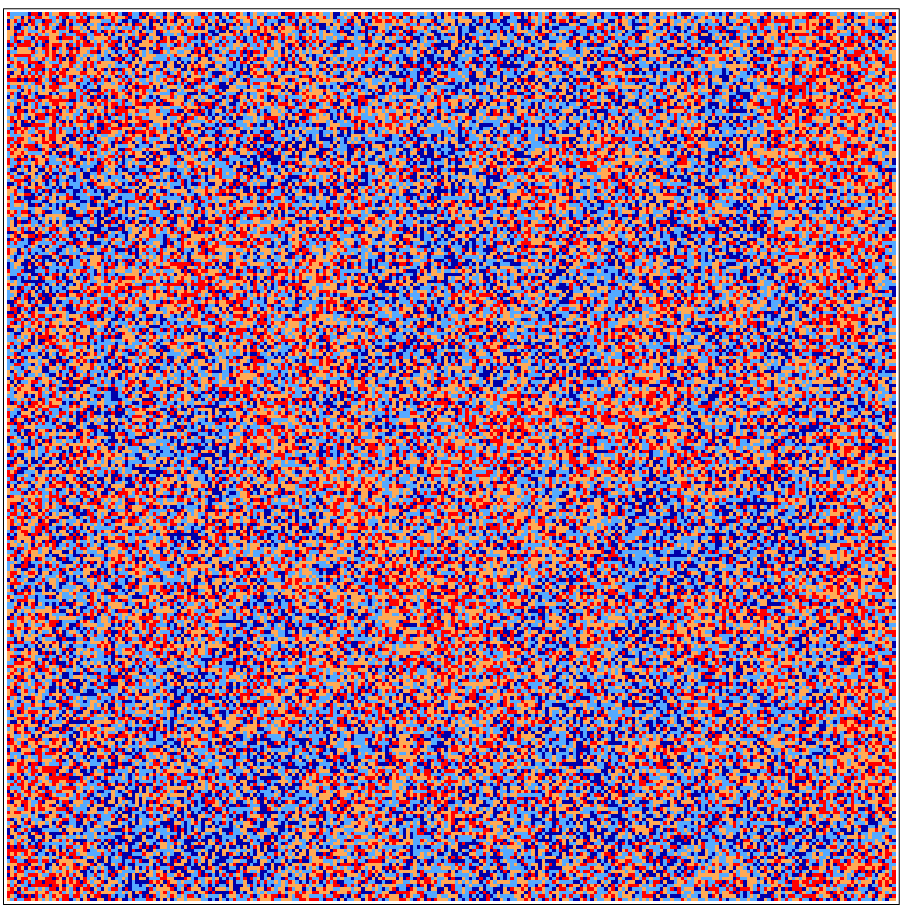}}
\put(0,4.05){\includegraphics[scale=.33]{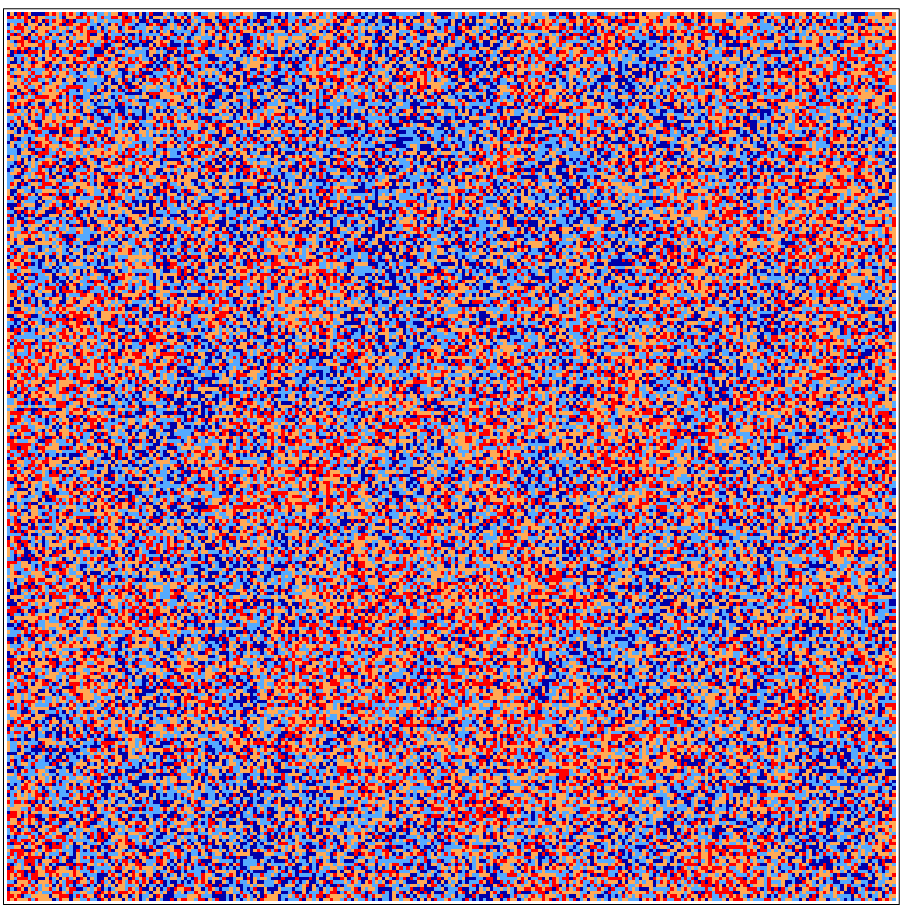}}
\put(1,4.05){\includegraphics[scale=.33]{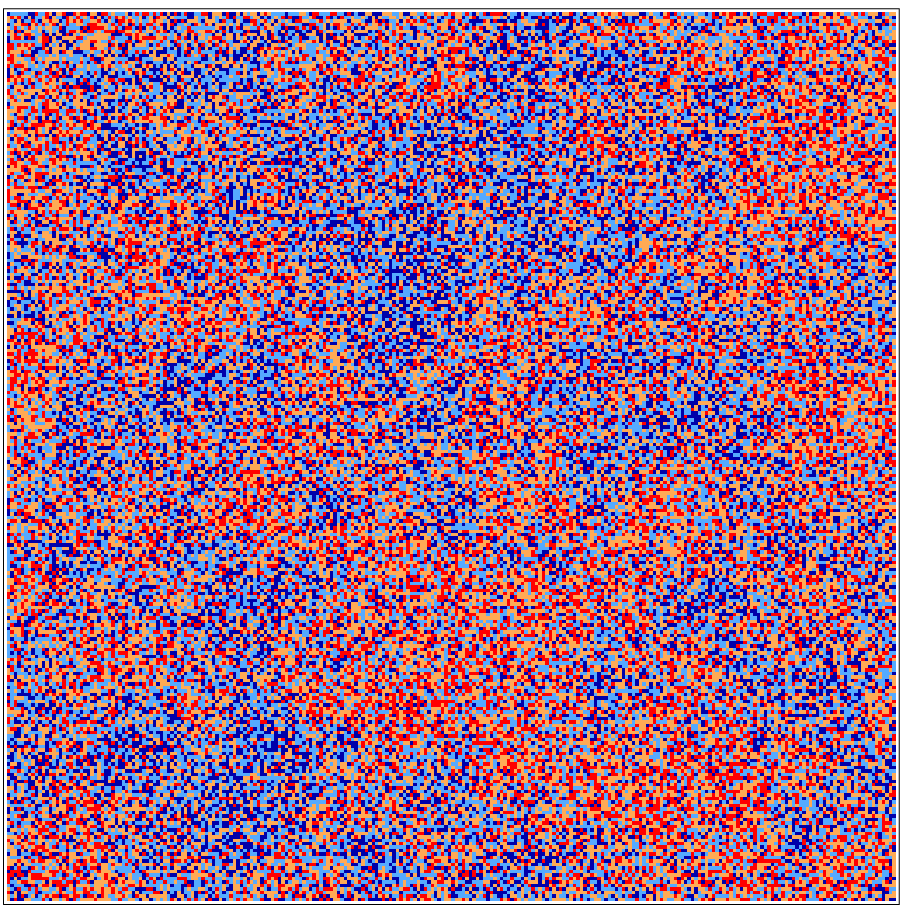}}
\put(2,4.05){\includegraphics[scale=.33]{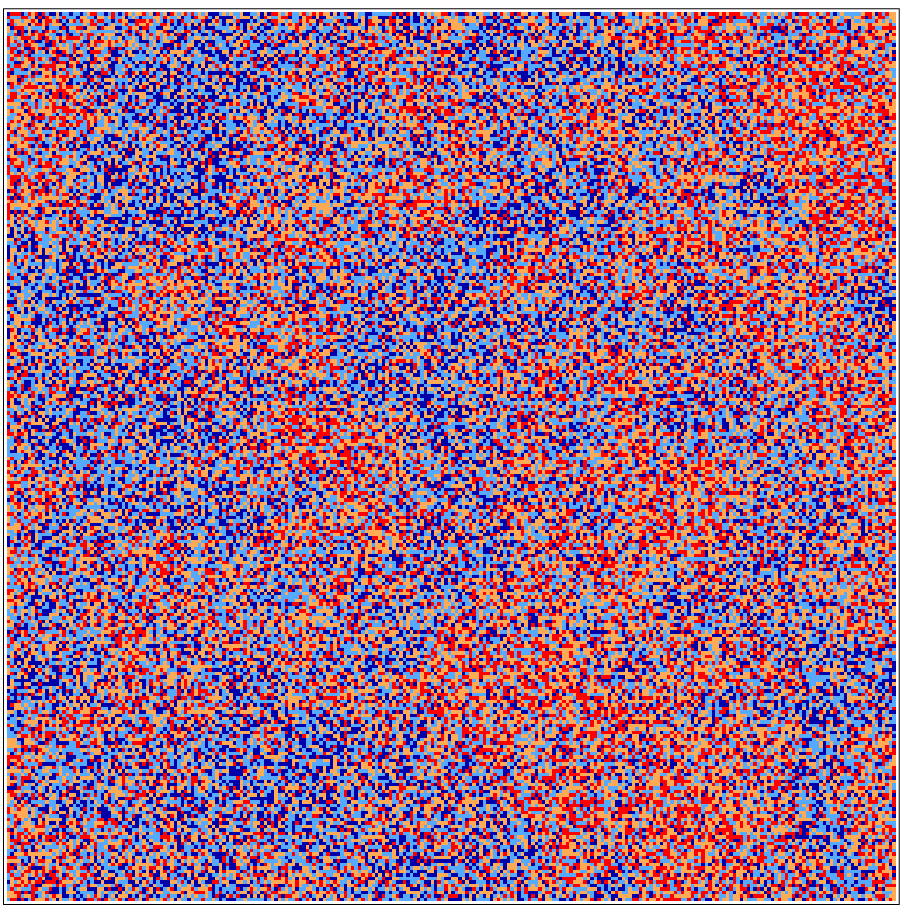}}

\put(-1,3.05){\includegraphics[scale=.33]{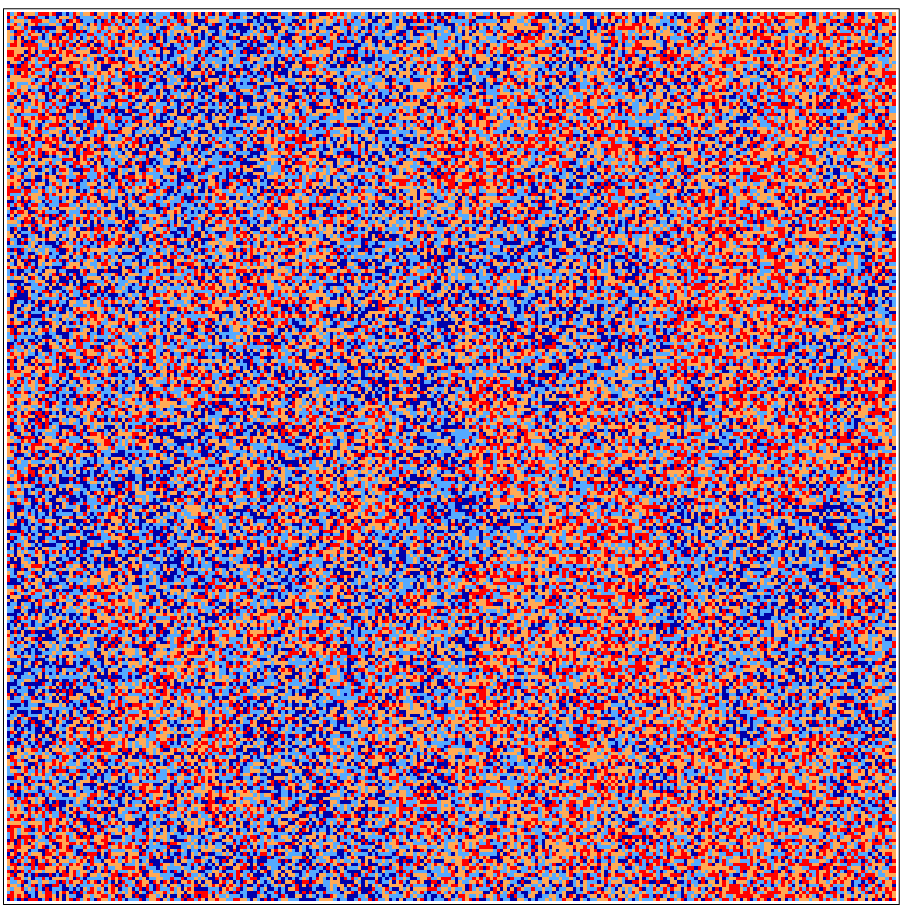}}
\put(0,3.05){\includegraphics[scale=.33]{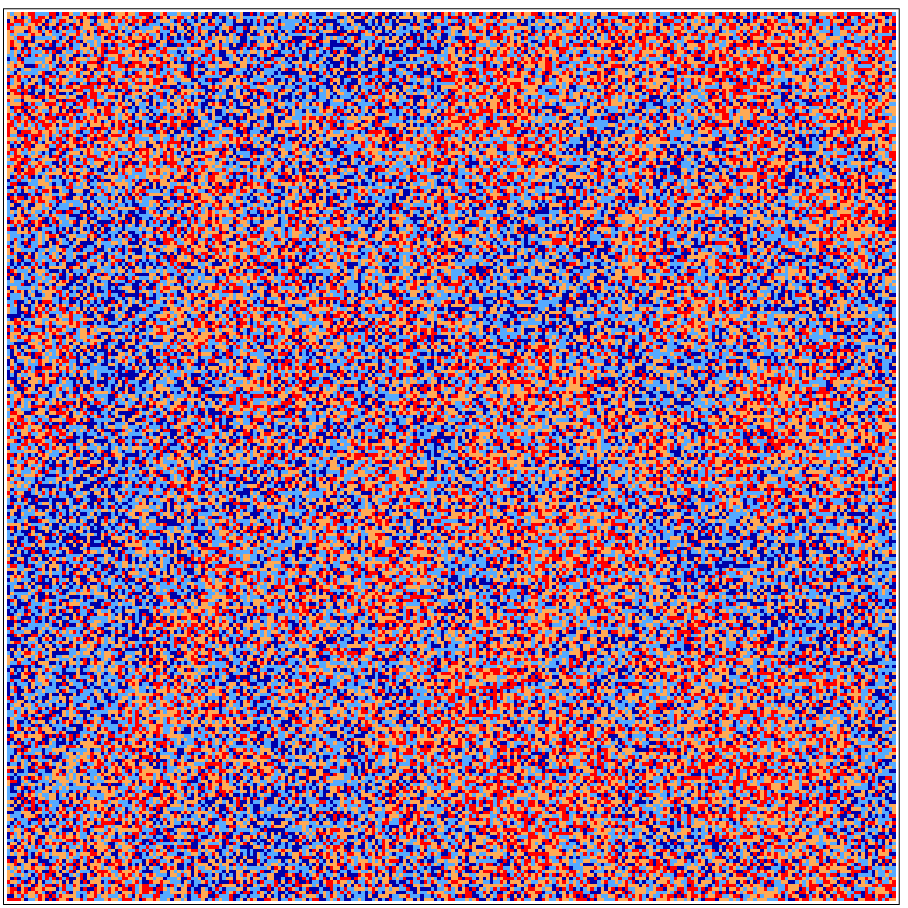}}
\put(1,3.05){\includegraphics[scale=.33]{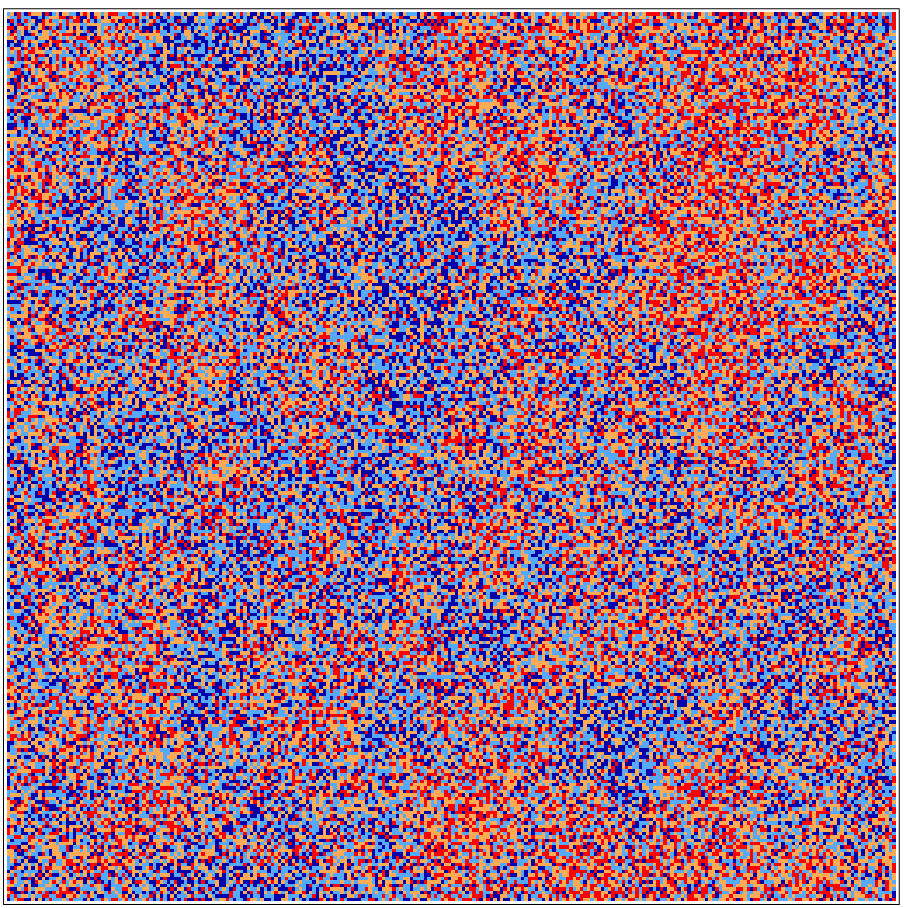}}
\put(2,3.05){\includegraphics[scale=.33]{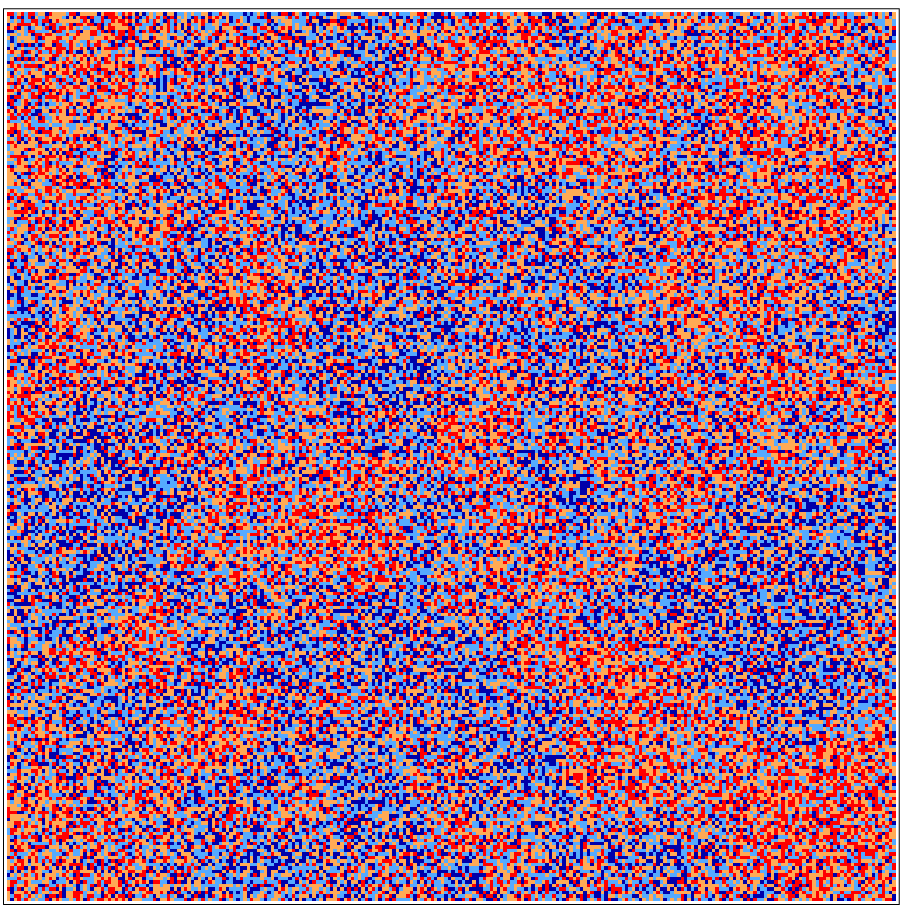}}
\end{picture}
%
%
%
\]
\caption{\label{fig:data1t}Time evolution of the Markov Chain
  described in Section \ref{ssec:dyntorus}. Here $L=256$, and times
  shown are $t=L^2,4L^2,9L^2,\ldots,144L^2$. Color code:
  $(0,1,2,3)=(\textrm{red},\textrm{orange},\textrm{cyan},\textrm{blue})$.}
%
\[
\setlength{\unitlength}{89pt}
\begin{picture}(1.964,3)
\put(-1,2){\includegraphics[scale=.33]{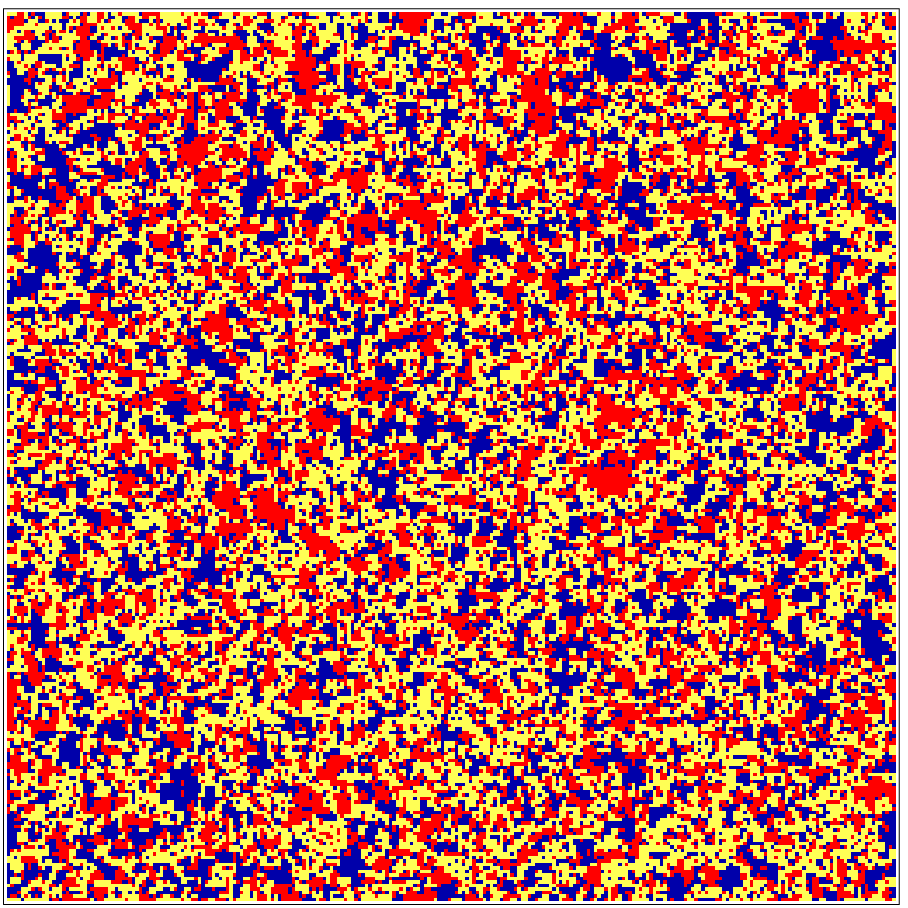}}
\put(0,2){\includegraphics[scale=.33]{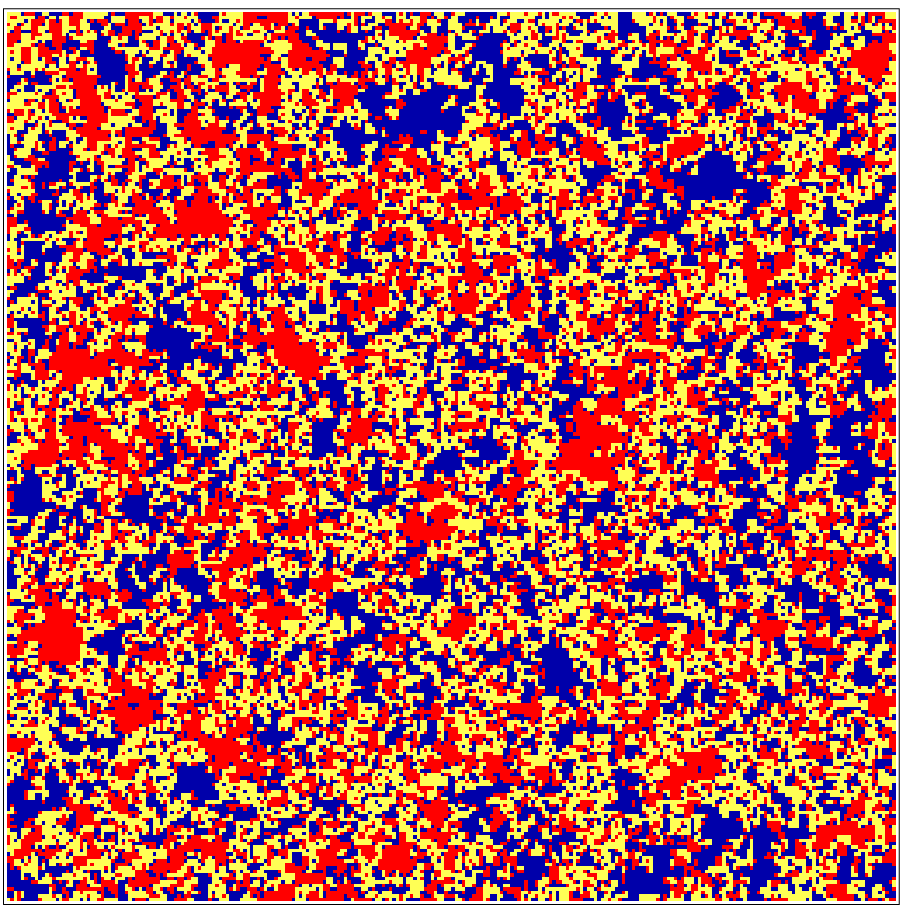}}
\put(1,2){\includegraphics[scale=.33]{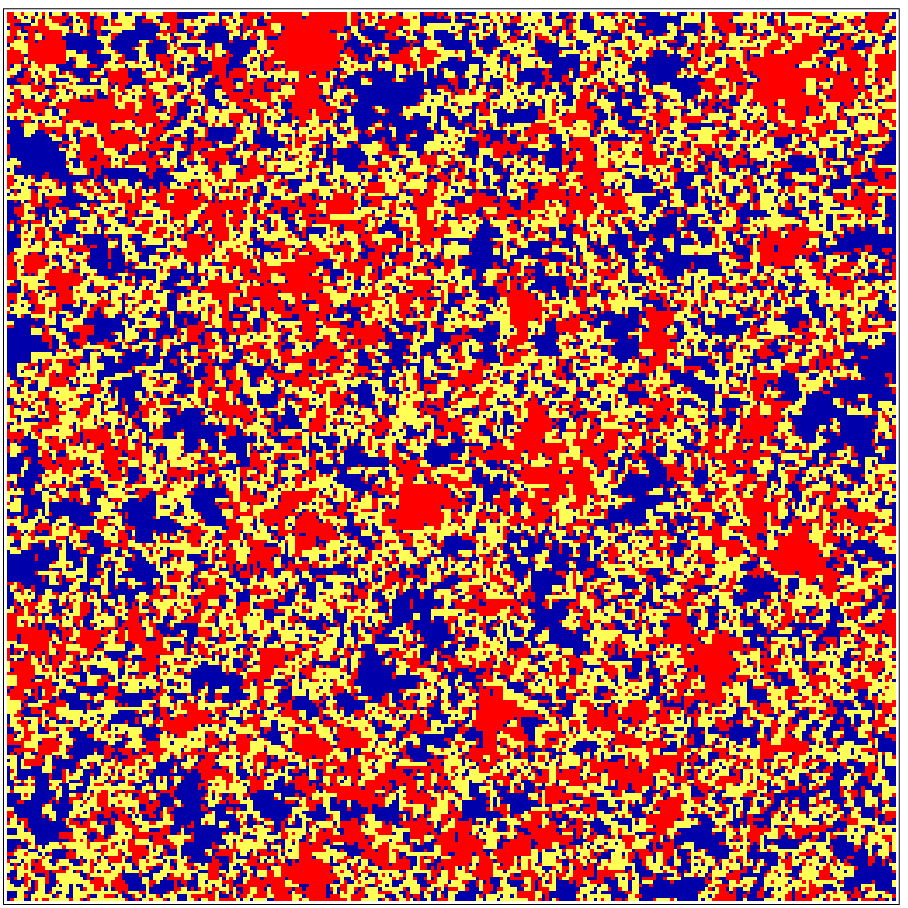}}
\put(2,2){\includegraphics[scale=.33]{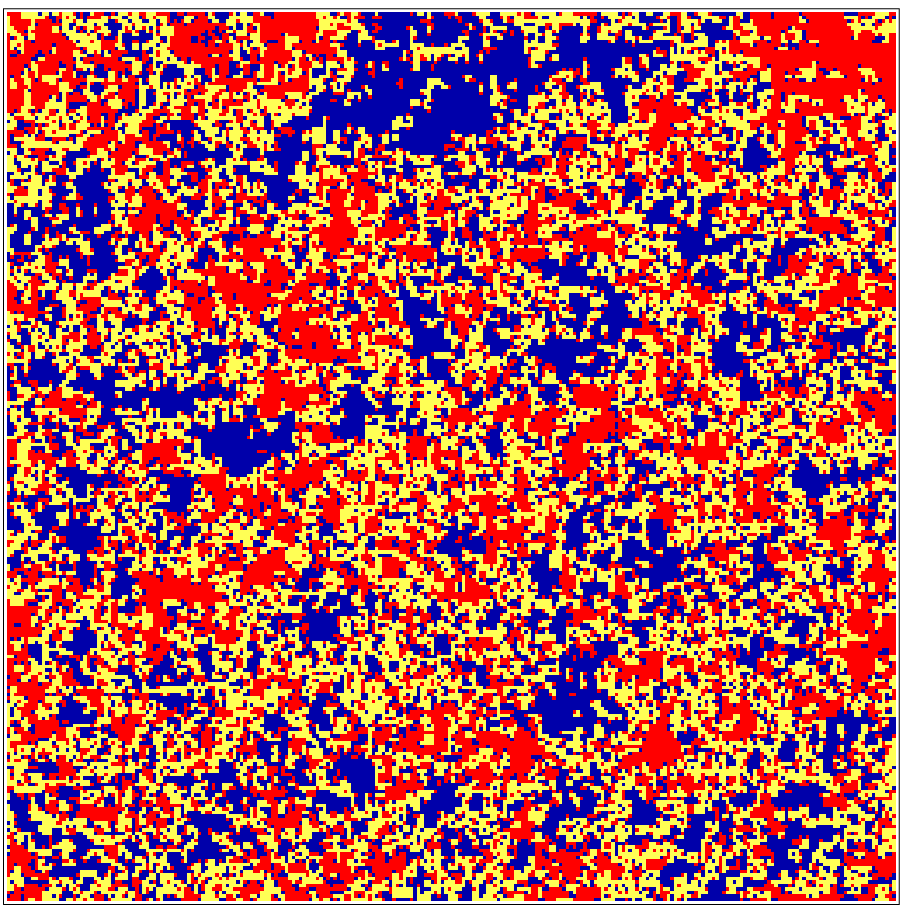}}

\put(-1,1){\includegraphics[scale=.33]{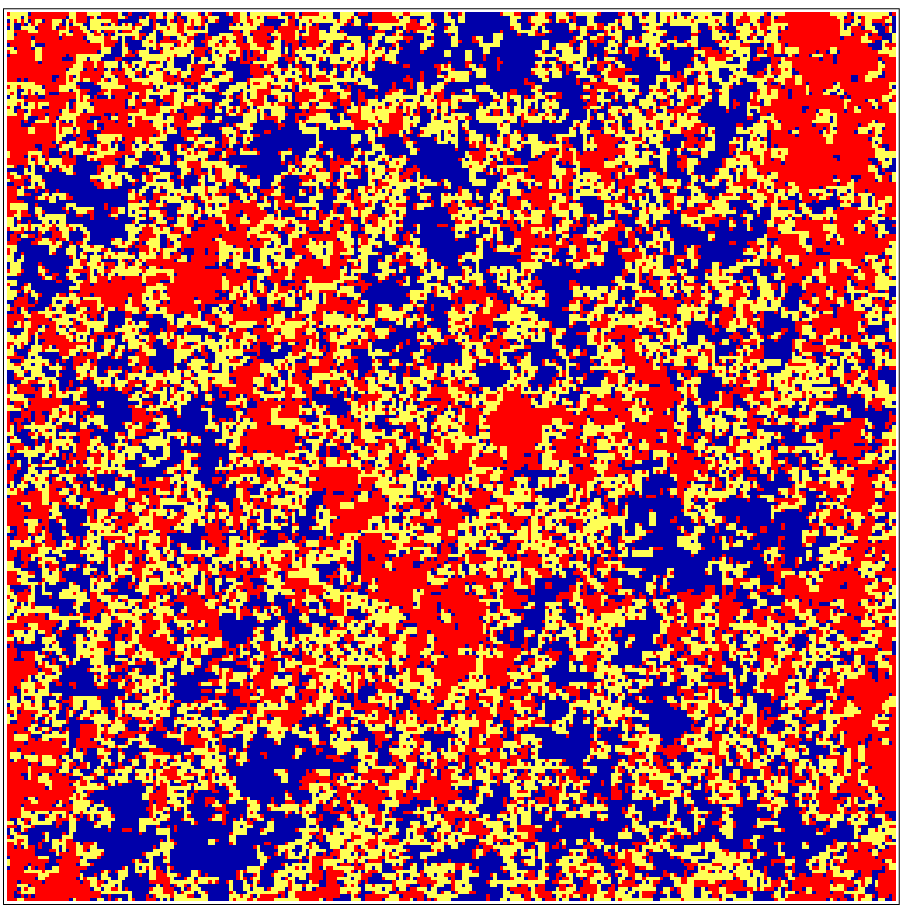}}
\put(0,1){\includegraphics[scale=.33]{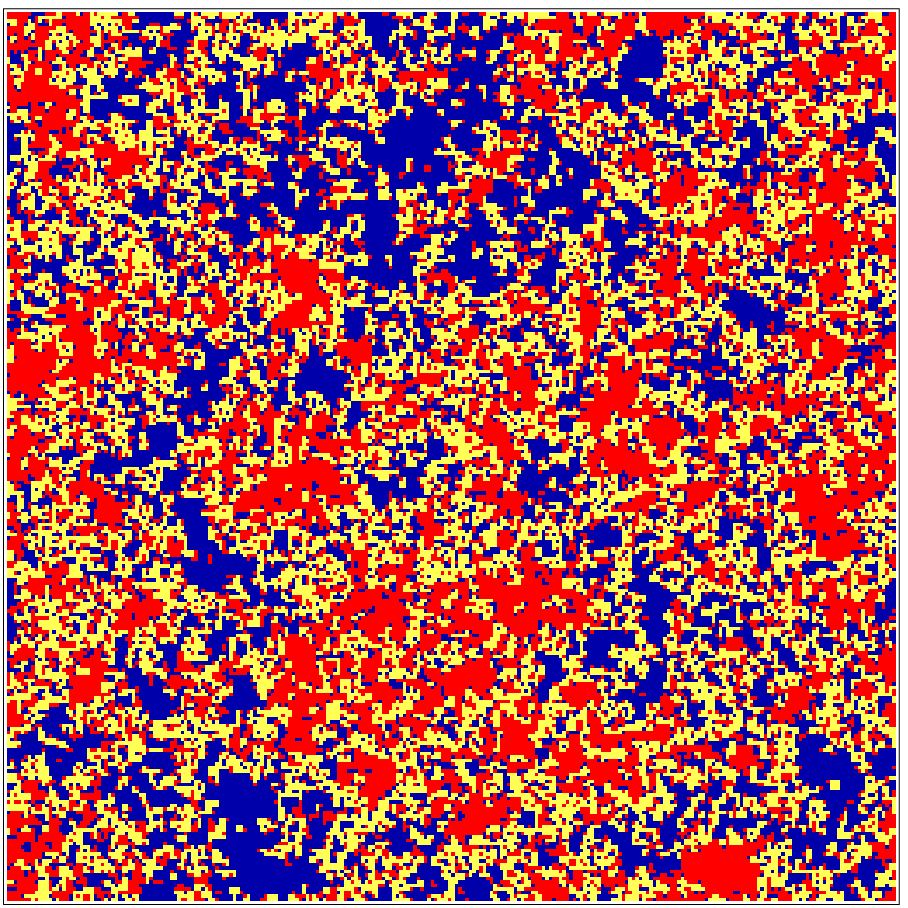}}
\put(1,1){\includegraphics[scale=.33]{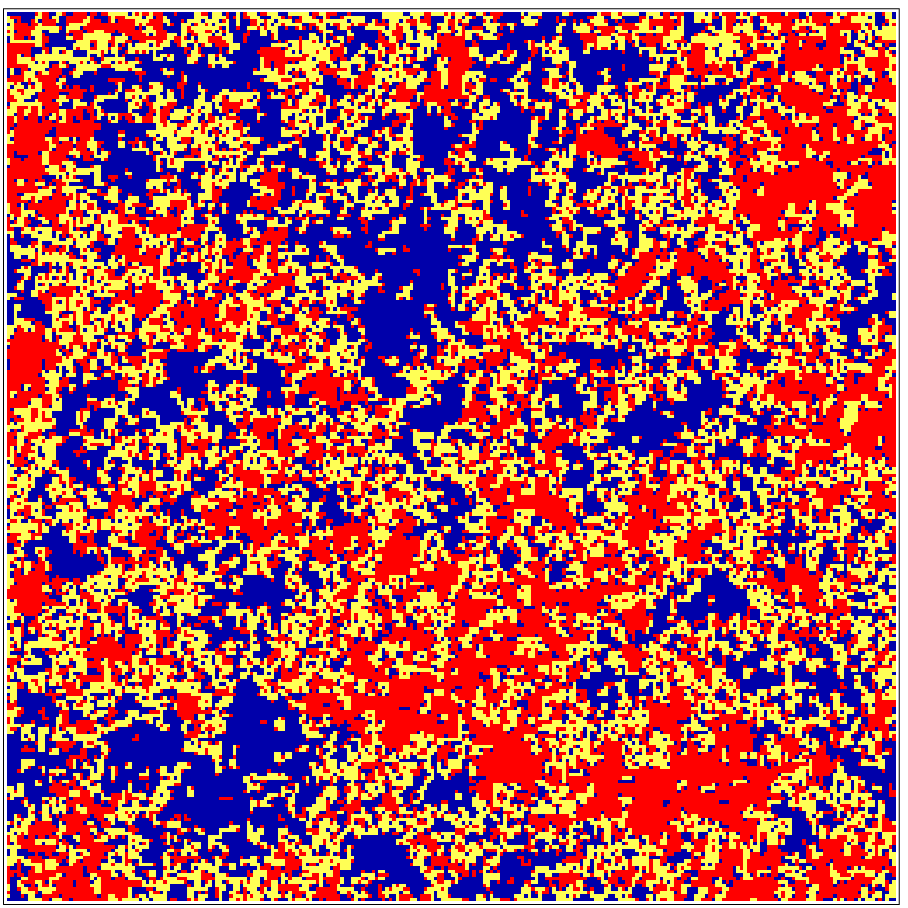}}
\put(2,1){\includegraphics[scale=.33]{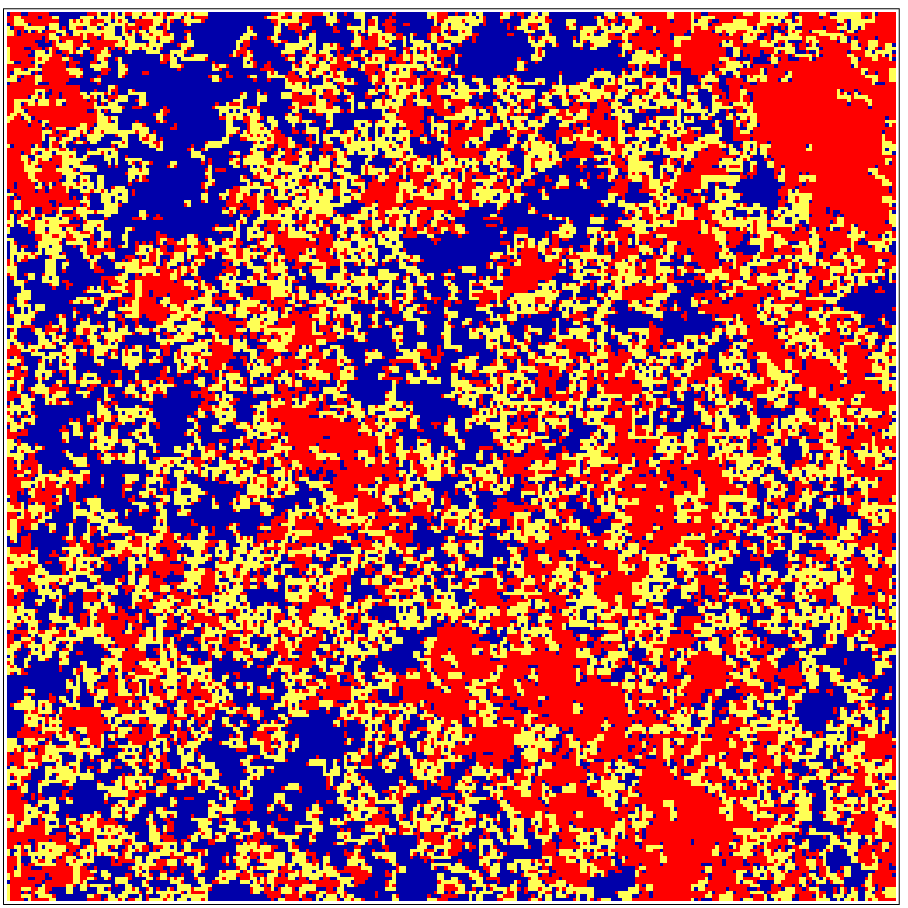}}

\put(-1,0){\includegraphics[scale=.33]{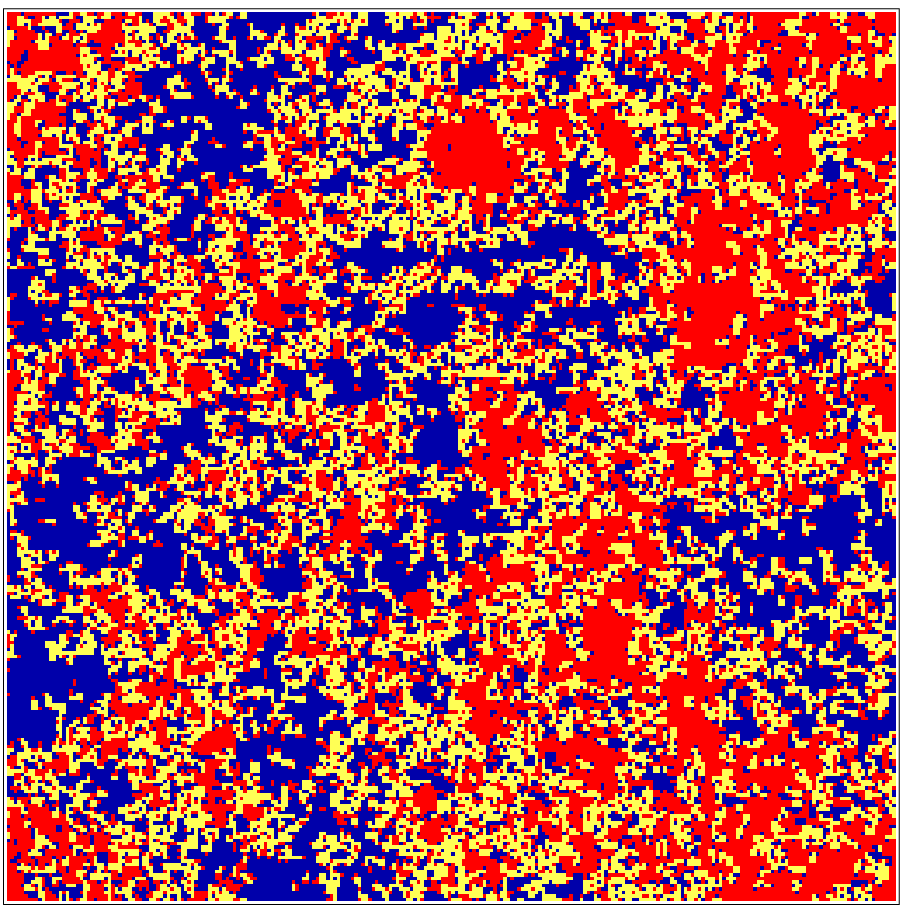}}
\put(0,0){\includegraphics[scale=.33]{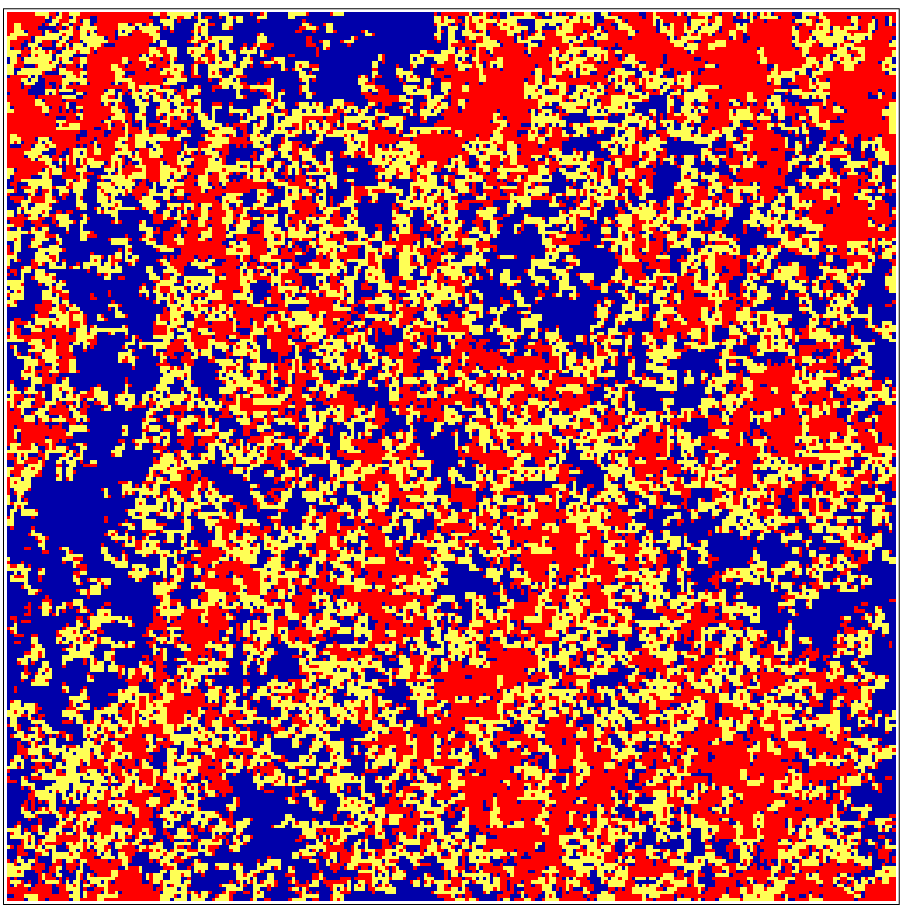}}
\put(1,0){\includegraphics[scale=.33]{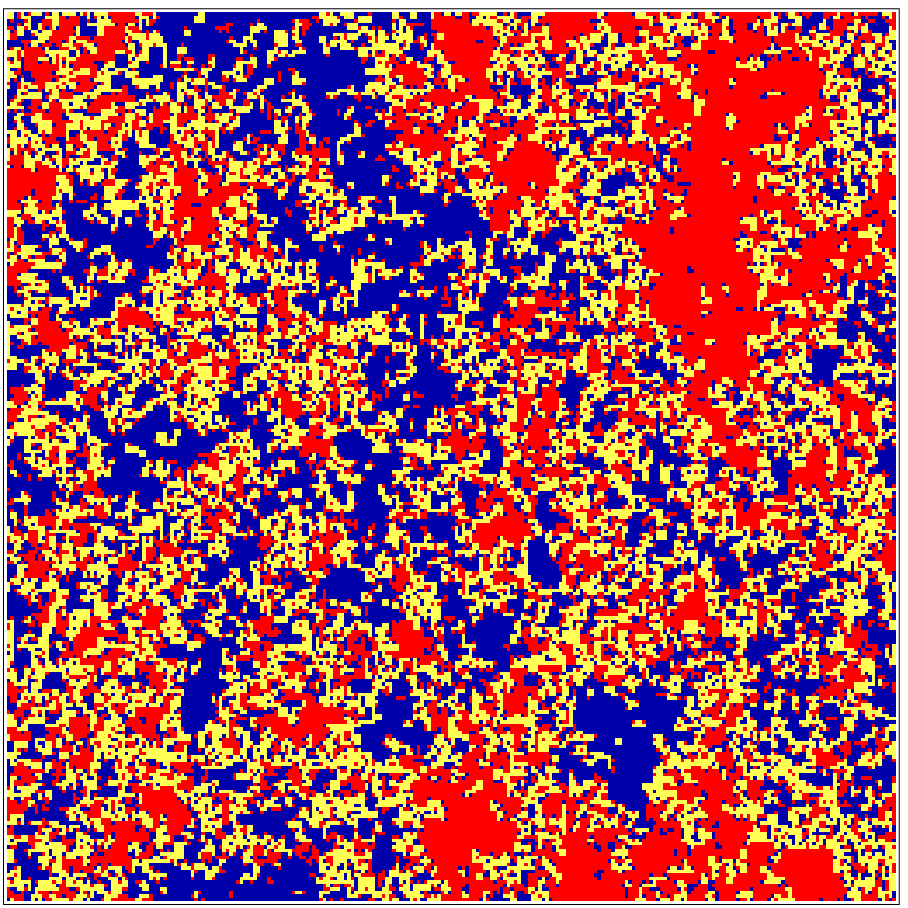}}
\put(2,0){\includegraphics[scale=.33]{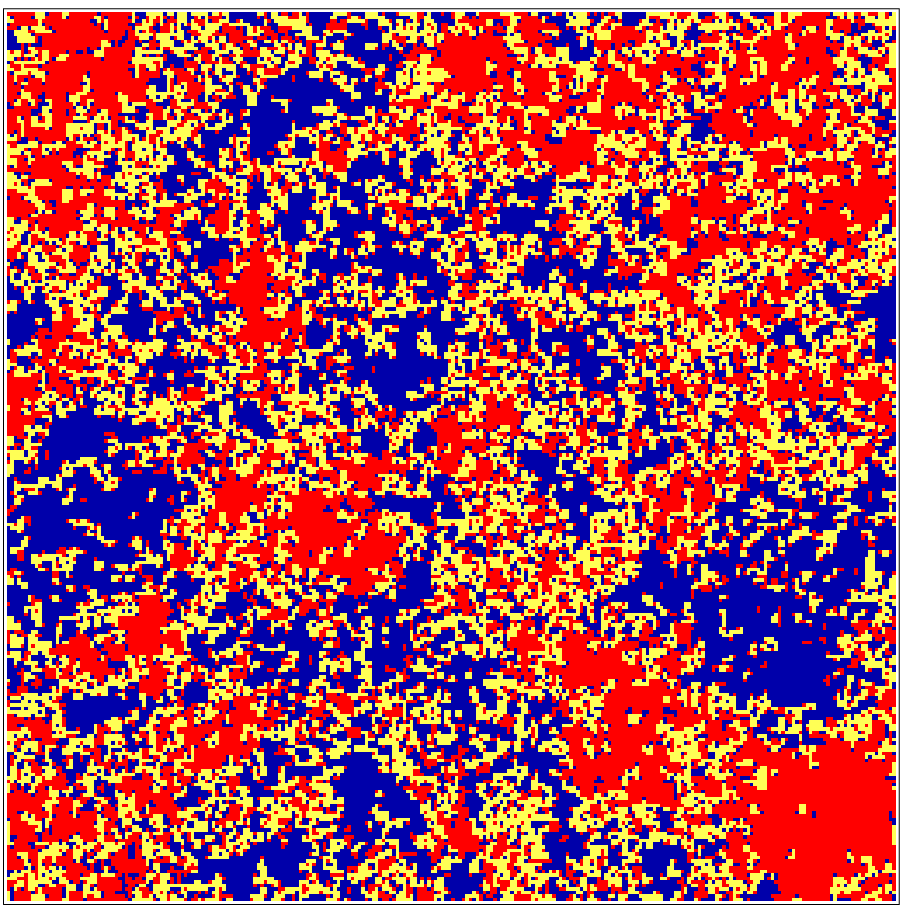}}
\end{picture}
%
%
\]
\caption{\label{fig:data1b} The  local parameter 
  associated to burning-test defined in the text of configurations in Fig.~\ref{fig:data1t}.}
\end{figure}

The precise time evolution is as follows. At each time, one of the
$2n^2$ operators $\{ a^\dg_j a^\fdg_i, a^\fdg_i a^\dg_j \}$ is chosen
uniformly at random, and applied to the previous state. For the
examples shown in Fig.~\ref{fig:data1t}, we have chosen as
initialisation the one best approximating a uniform distribution of
the mass, $z_{(x,y)} = 1$ if $x+y$ is odd and $2$ if it is even. As
seen from the figure,
the coarsening is already evident at the level of the height
variables.  It is however much more evident at the level of a local
parameter suggested by the Burning Test. The ordinary Burning
Test cannot be performed here, as we have no boundary. We choose as
{\em effective boundary} the set of sites with maximal height. More
precisely, we record the sites that perform at least one toppling, in
the avalanche following the replacement $3 \to 4$, and colour them in
blue.  Similarly, we record the sites that perform at least one
anti-toppling, in the avalanche following the replacement $0 \to -1$,
and colour them in red. In case both a toppling and an antitoppling
have occurred, we break the tie by observing the initial height. In
case neither a toppling nor an antitoppling have occurred, we use a
{\em neutral} light-yellow colour. Thus, pictorially, blue, red and yellow
regions correspond to regions which are recurrent, anti-recurrent, and
simultaneously transient and anti-transient.  The resulting analysis
of the configurations is shown in Fig.~\ref{fig:data1b}.

\subsection{Dynamics with idempotent operators}
\label{ssec:dynproj}

\begin{figure}[t]
\[
\setlength{\unitlength}{93pt}
\begin{picture}(1.67,3.05)
\put(-.97,2){\includegraphics[scale=.85]{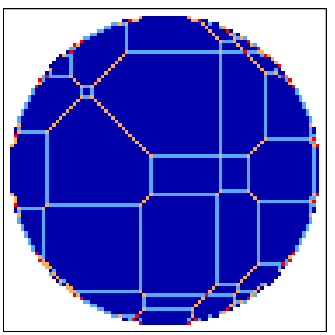}}
\put(0,2){\includegraphics[scale=.85]{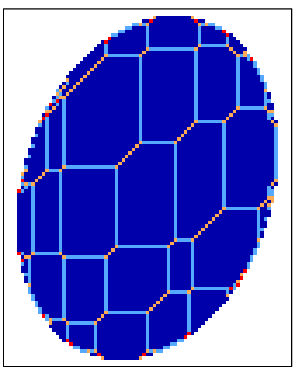}}
\put(.89,2){\includegraphics[scale=.85]{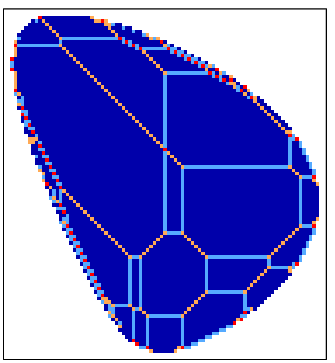}}
\put(1.87,2){\includegraphics[scale=.85]{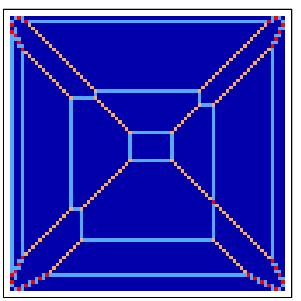}}

\put(-.97,1){\includegraphics[scale=.85]{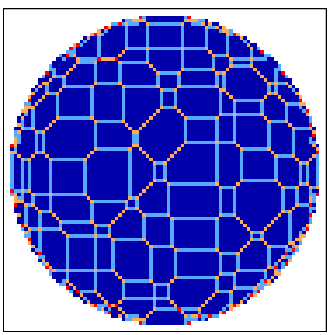}}
\put(0,1){\includegraphics[scale=.85]{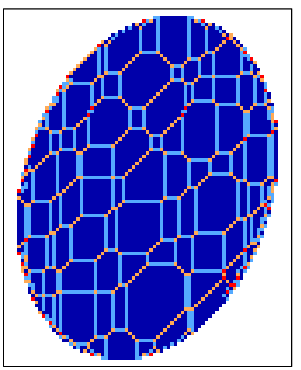}}
\put(.89,1){\includegraphics[scale=.85]{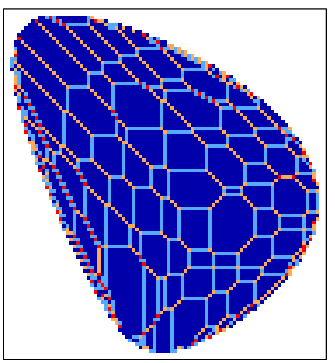}}
\put(1.87,1){\includegraphics[scale=.85]{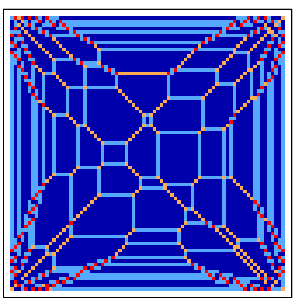}}

\put(-.97,0){\includegraphics[scale=.85]{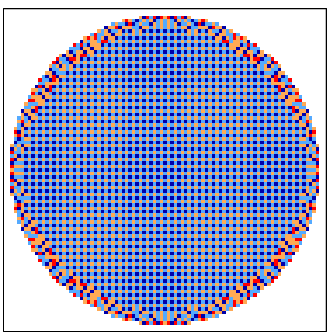}}
\put(0,0){\includegraphics[scale=.85]{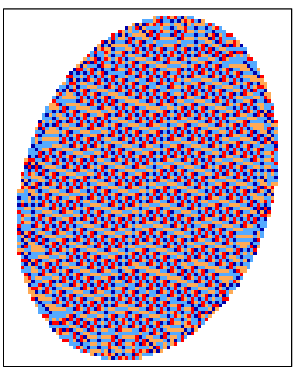}}
\put(.89,0){\includegraphics[scale=.85]{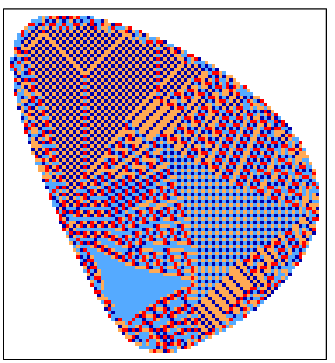}}
\put(1.87,0){\includegraphics[scale=.85]{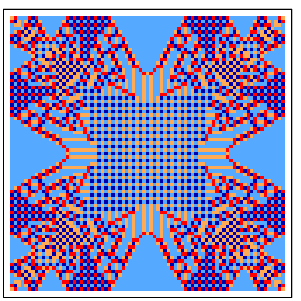}}
\end{picture}
%
%
\]
\caption{\label{fig.aat} Configurations obtained with the
procedure described in Section \ref{ssec:dynproj}. Top, middle and
bottom row correspond to $t=32$, $t=1024$, and to the fixed point of
the dynamics. The columns show different domains. From left to right:
a circle; an ellipse with axes rotated by $\arctan 1/3$ w.r.t.\ the
cartesian axes, and ratio 2 between height and width; a typical
algebraic curve of degree 3, more precisely 
$4 x^2 + 4 y^2 + 3 x y + 2 x^2 y + x^3 = 7$; a square.}
\end{figure}

Our second dynamics starts from the maximally filled configuration,
$z_i=3$ for all $i$, and acts with the idempotent combinations
$a^\dg_i a^\fdg_i$ at randomly-chosen sites. 

As we know from Theorem \ref{theo.orchidweak}, this dynamics is
absorbent on a unique configuration, identified with the
multitoppling relaxation of the initial configuration, where pairs of
adjacent sites both with height 3 are unstable. 

Again, interesting features emerge at short times, when the
configuration takes the form of a {\em web of strings}, satisfying a
classification theorem and a collection of incidence rules~\cite{noi-epl}.

On the other extremum of the dynamics, at the fixed point we have
configurations showing remarkable regularities, in the form of
{\em patches}, that is, a local two-dimensional periodicity on portions of
the domain~\cite{Ostojic}. When
more patches are present, they follow an incidence rule~\cite{DSC}.

If the initial domain is an elliptic portion of the square lattice, a
specially higher regularity emerges.  Say that the linear dimension of
the domain is of order $L$, and the slope of the symmetry axes is a
{\em small} rational $p/q$ (with both $p$ and $q$ of $\mathcal{O}(1)$ in
$L$). Then, in the limit $L \to \infty$, we observe the emergence of a
very simple structure of patches and strings: we have a unique patch,
crossed by strings of a unique type, parallel to one of the two
symmetry axes. This fact is in agreement with the general theory
developed in \cite{Ostojic,DSC,noi-epl}, as the toppling vector at a
coarsened level is a quadratic form in the coordinates $x$ and $y$,
that should vanish at the boundary of the domain, and the contour
lines of quadratic forms are conics, i.e.\ plane algebraic curves of
degree~2.

In Fig.~\ref{fig.aat} we present configurations obtained with the
procedure described above, starting with the maximally-filled
configuration, $z_i=3$ for all $i$, on portions of the square lattice
of various shapes, in order to highlight the features outlined above:
a disk, an ellipse, a smooth domain which is not a conic (it is an
algebraic curve of degree 3), and a square.

\subsection{A simple deterministic dynamics}
\label{ssec:dyntapis}

\newsavebox{\sma}
\savebox{\sma}{$\big(\begin{smallmatrix} 2 & 1 \\ 1 & 2 \end{smallmatrix} \big)$}
\newsavebox{\smb}
\savebox{\smb}{$(\begin{smallmatrix} 2 \end{smallmatrix})$}
\newsavebox{\smc}
\savebox{\smc}{$\big(\begin{smallmatrix} 2 & 2 \\ 2 & 1 \end{smallmatrix} \big)$}
\newsavebox{\smd}
\savebox{\smd}{$\Big(\begin{smallmatrix} 3 & 2 & 2 \\ 2 & 2 & 2 \\ 2 & 2 &
    2 \end{smallmatrix}\Big)$}

\begin{figure}[ht]
\[
\setlength{\unitlength}{83pt}
\begin{picture}(2.,4.1)(0,-1)
\put(-1,2.15){\includegraphics[scale=.65]{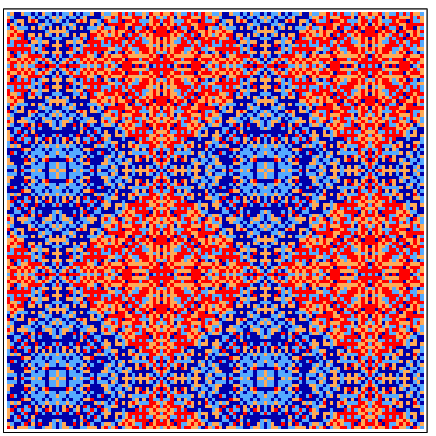}}
\put(0,2.15){\includegraphics[scale=.65]{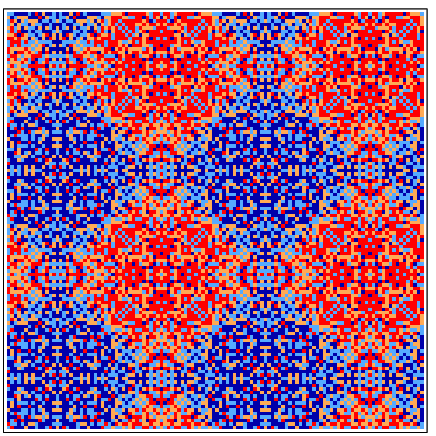}}
\put(1,2.15){\includegraphics[scale=.65]{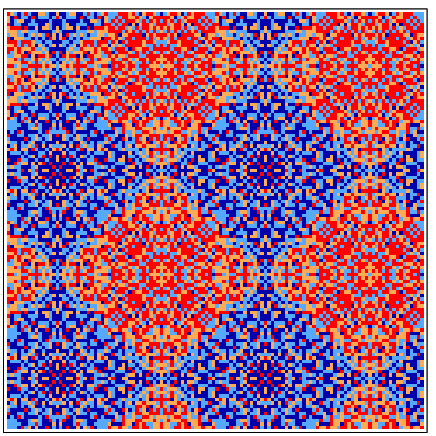}}
\put(2.05,2.15){\includegraphics[scale=.65]{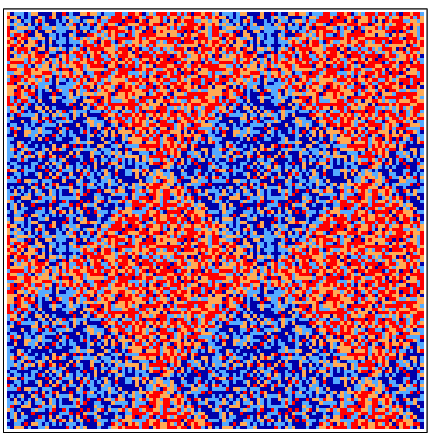}}

\put(-1,1.1){\includegraphics[scale=.65]{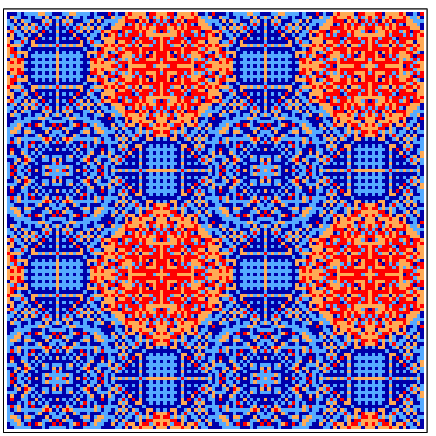}}
\put(0,1.1){\includegraphics[scale=.65]{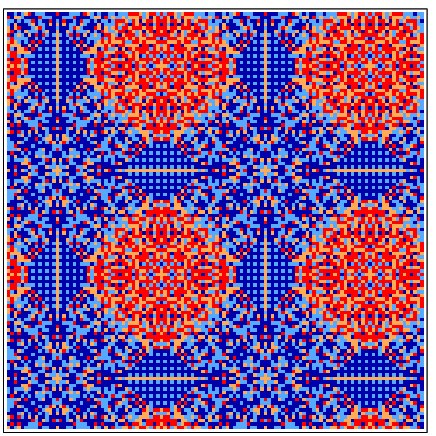}}
\put(1,1.1){\includegraphics[scale=.65]{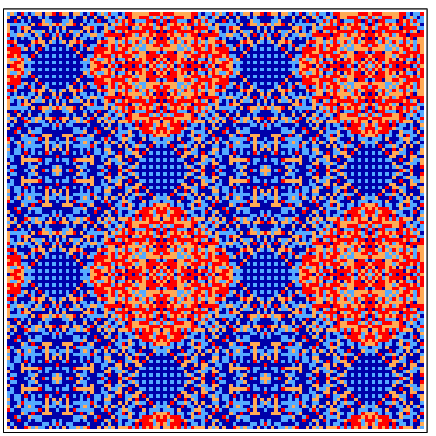}}
\put(2.05,1.1){\includegraphics[scale=.65]{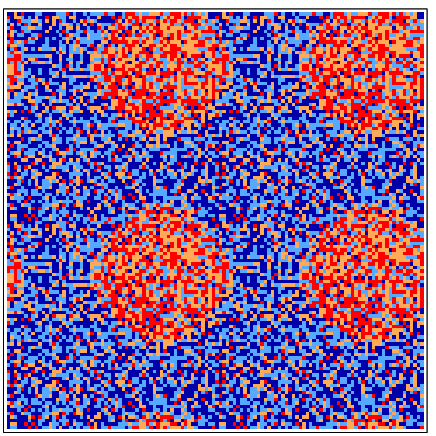}}

\put(-1,0.05){\includegraphics[scale=.65]{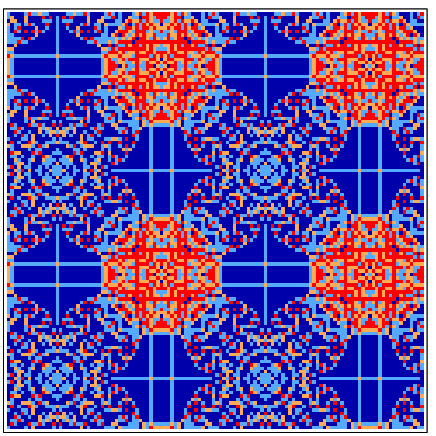}}
\put(0,0.05){\includegraphics[scale=.65]{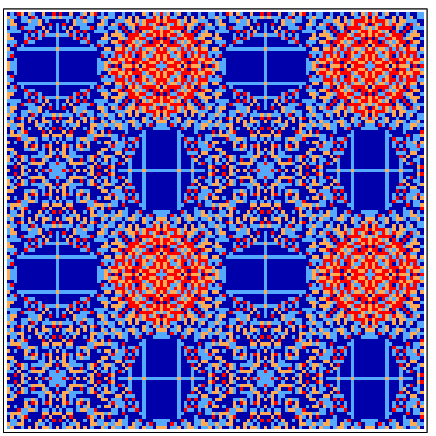}}
\put(1,0.05){\includegraphics[scale=.65]{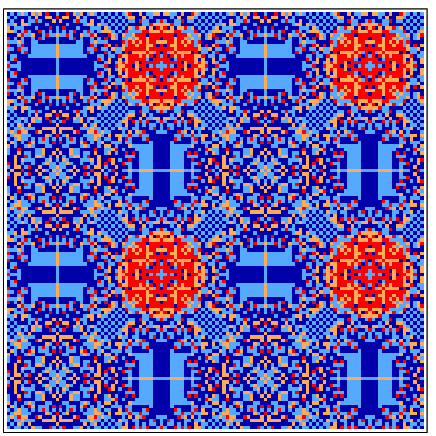}}
\put(2.05,0.05){\includegraphics[scale=.65]{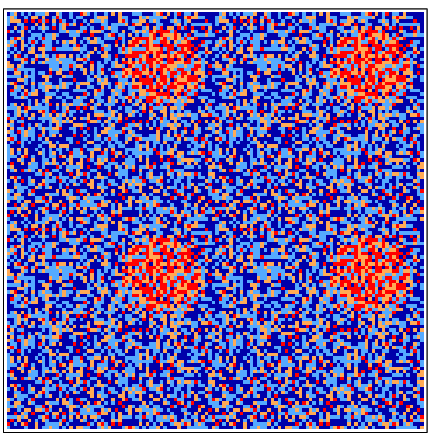}}

\put(-1,-1){\includegraphics[scale=.65]{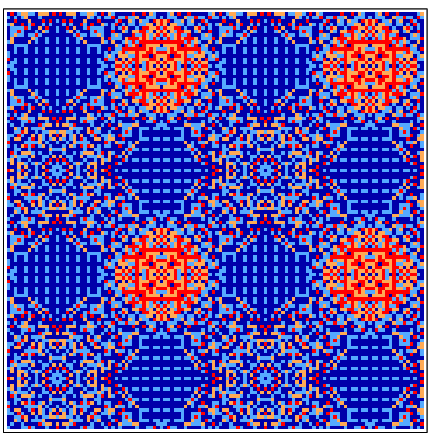}}
\put(0,-1){\includegraphics[scale=.65]{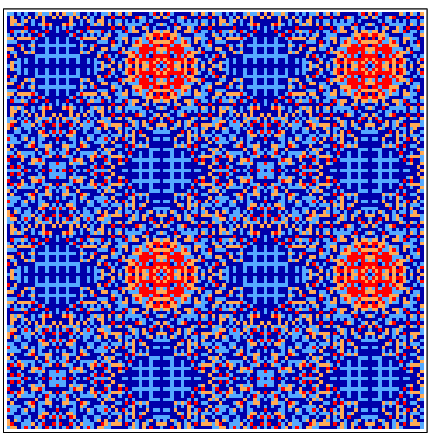}}
\put(1,-1){\includegraphics[scale=.65]{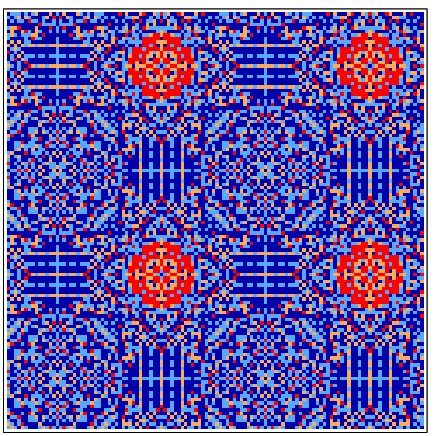}}
\put(2.05,-1){\includegraphics[scale=.65]{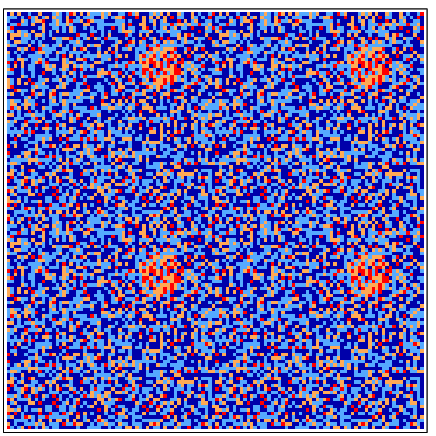}}
\end{picture}
%
%
%
\]
\caption{\label{fig.tapis_ex}Examples of configurations, for $L=60$, observed in
  the dynamics described in Section~\ref{ssec:dyntapis}.   The domain is repeated periodically $2 \times 2$
  times, in order to improve visualisation.
 The different rows correspond to  densities $\rho=1.5$, 1.75,
  2, 2.111$\ldots$.
  The three left-most columns show, respectively, configurations at time $t=3000$, 6000 and 9000, obtained by
  starting from periodic configurations prepared by using the tiles \usebox{\sma},  \usebox{\smc}, \usebox{\smb},   and \usebox{\smd}. The last column shows, for comparison, the configurations at $t=3000$ obtained  by starting from random configurations at the given densities.
}
\end{figure}

Our third and last example of dynamics is deterministic, and
corresponds to the sand flow from a unique source site $i$ to a
unique sink site $j$ in the lattice, through the iterated application
of the operator $a^\dg_j a^\fdg_i$. 

Besides producing beautiful drawings, this dynamics has an interesting
phenomenological feature.  Contrarily to the na\"ive intuition of a
smooth hydrodynamic flow from the source to the sink, in typical
configurations at the steady state, the density profile presents a
{\em shock}, between a spatial region analogous to ordinary BTW and a
region analogous to the involution image of the BTW, both regions
presenting large avalanches (of the appropriate type). 

This phenomenology seems an analogue of the shock profiles observed in
various non-equilibrium systems, most notably a two-dimensional
analogue of a feature of the TASEP \cite{DEPH}, an exactly-solvable
model in one dimension, in the case in which two species of particles
are considered~\cite{DerridShocks}.  

We note that, in comparison with the previous realisations, this
dynamics presents short thermalisation times and very moderate aging.
(both the thermalisation time and the aging effect seem to become more
relevant as $\rho$ approaches $\rho^*$).  A partial justification is
the fact that this dynamics is deterministic and thus, after a
transient, it follows a periodic orbit, although the growth of the
period with the system size is extremely fast. Recall that the length
of the orbit of the operator $a_j^{-1} a_i$, acting in the torus $G
\cong \mathbb{Z}_{d_1} \times \mathbb{Z}_{d_2} \times \cdots \times
\mathbb{Z}_{d_g}$ of recurrent configurations, for the abelian version
of the model, is a divisor of the length of the orbit of $a^\dg_j
a^\fdg_i$ acting on $S$, due to the fact that the equivalence classes
are preserved in the framework with both topplings and antitopplings.

We presents configurations realised on a $L \times L$ torus, with sink
and source at antipodal points, i.e.\ acting with the operator
$a^\dg_j a^\fdg_i = a^\dg_{(L/2,L/2)} a_{(0,0)}$, which is the
simplest and most symmetric realisation of the ideas above. We studied
both the case in which the initial state is periodic, and the case of
random initialisation with given prescribed density.  The
configurations are presented in Fig.~\ref{fig.tapis_ex}.


%
%


\end{document}